\documentclass[]{aa}

\usepackage{natbib}
\usepackage{graphicx}
\usepackage{color}
\usepackage{amsmath}
\usepackage{url}

\newcommand\bb[1] {   \mbox{\boldmath{$#1$}}  }
\newcommand\del{\bb{\nabla}}
\newcommand\bcdot{\bb{\cdot}}
\newcommand\btimes{\bb{\times}}
\newcommand{\mean}[1]{\langle #1 \rangle}

\begin{document}

\title{Thermodynamics of the dead-zone inner edge in protoplanetary disks}
\author{Julien Faure \inst{1}
\and S\'ebastien Fromang \inst{1} 
\and Henrik Latter \inst{2}}  

\offprints{J.Faure}

\institute{Laboratoire AIM, CEA/DSM - CNRS - Universit\'e Paris 7,
  Irfu/Service d’Astrophysique, CEA-Saclay, 91191 Gif-sur-Yvette, 
France \and Department of Applied Mathematics
and Theoretical Physics, University of Cambridge, Centre for
Mathematical Sciences, Wilberforce Road, Cambridge, CB3 0WA, UK\\
\email{julien.faure@cea.fr}}

\date{Accepted/ Received/ in original form}

\abstract{In protoplanetary disks, the inner boundary between the turbulent and laminar regions could be a
promising site for planet formation, thanks to the trapping of solids at the
boundary itself or in vortices generated
by the Rossby wave instability. 
%% Activity at this
%% radius will also influence the thermodynamic structure of much of the
%% dead-zone (including features such as the ice line). 
At the interface,
 the disk thermodynamics and the turbulent dynamics
 are entwined because of the importance of turbulent
 dissipation and thermal ionization.
 Numerical models of the boundary, however, have neglected
 the thermodynamics, and thus miss a part of the physics.}  
{The aim of this paper is to numerically investigate the interplay between
  thermodynamics and dynamics in the inner regions of protoplanetary
  disks by properly accounting for turbulent heating and the dependence of the
resistivity on the local temperature.}  
{Using the Godunov code RAMSES, we performed a series of 3D global numerical
simulations of protoplanetary disks in the cylindrical limit,
including turbulent heating and a simple prescription for radiative
cooling.}
{We find that waves excited by the turbulence
significantly heat the dead zone, and we subsequently provide a simple theoretical
framework for estimating the wave heating and consequent temperature
profile. In addition, our simulations reveal that the dead-zone inner
edge can propagate outward into the dead zone, before staling at a critical
radius that can be estimated from a mean-field model. The engine
driving the propagation is in fact density wave heating close to the
interface. A pressure maximum appears at the interface in all
simulations, and we note the emergence of the Rossby wave instability in
simulations with extended azimuth.}
{Our simulations illustrate the complex interplay between
  thermodynamics and turbulent dynamics in the inner regions of
  protoplanetary disks. They also reveal how important activity at
  the dead-zone interface can be for the dead-zone thermodynamic structure.} 


\maketitle

\authorrunning{J.Faure et al.}
\titlerunning{Dead zone inner edge dynamics}

\section{Introduction}

Current models of protoplanetary (PP) disks are predicated on the idea that
significant regions of the disk are too poorly ionized to sustain MRI
turbulence. These PP disks are thought to comprise a turbulent body of
plasma (the `active zone') enveloping a region of cold quiescent gas,
in which accretion is actually absent (the `dead zone') (Gammie
1996, Armitage 2011). These models posit a critical inner radius ($\sim
1$ au) within which the disk is fully turbulent and beyond which the
disk exhibits turbulence only in its surface layers, for a range of
radii ($1-10$ au)
(but see Bai \& Stone 2013 for a complication of this picture).  

The inner boundary between the MRI-active and dead regions is crucial
for several key processes. Because there is a mismatch in accretion
across the boundary, a pressure maximum will naturally form at this
location which (a)
may halt the inward spiral of centimetre to metre-sized planetesimals \citep{kretkeetal09}, 
and (b) may excite a large-scale vortex instability (`Rossby wave
instability') \citep{1999ApJ...513..805L} that may promote dust accumulation, hence
planet formation
\citep{barge&sommeria95,2009A&A...497..869L,meheutetal12}. 
On the other hand, the interface
will influence the radial profiles of the dead zone's
thermodynamic variables ---
temperature and entropy, most of all.
Not only will it affect the global disk structure and key disk features
(such as the ice line), but
 the interface will also control the preconditions for dead-zone instabilities that feed
 on the disk's small adverse entropy gradient, such as the subcritical
 baroclinic instability and double-diffusive instability
 \citep{lesur&pap10,2010MNRAS.405.1831L}.

Most studies of the interface have been limited to isothermality. This
is a problematic assumption because of the pervasive interpenetration of dynamics
and thermodynamics in this region, especially at the midplane. 
Temperature depends on the
turbulence via the dissipation of its kinetic and magnetic fluctuations, but
the MRI turbulence, in turn, depends on the temperature through the
ionization fraction, which is determined by thermal ionization
 \citep{pneuman&mitchell65,umebayashi&nakano88}.
Because of this feedback loop, the temperature is not an additional piece of
physics that we add to simply complete the picture; it is instead at the heart
of how the interface and its surrounding region work. 
One immediate consequence of this feedback is that much of the midplane
gas inward of 1 au is bistable: if the gas at a certain radius begins as
cold and poorly ionized, it will remain so; conversely, if it begins
hot and turbulent, it can sustain this state via its own waste heat
\citep{latter&balbus12}. Thus two stable states are available at any given 
radius in the bistable region. This complicates the question of where
the actual location of the dead zone boundary lies. It also
raises the possibility that the boundary is not static, and may not
even be well defined. Similar models
 have also been explored in the context of FU Ori outbursts
\citet{2010ApJ...713.1134Z,2009ApJ...694.1045Z,2009ApJ...701..620Z}.

In this paper we simulate these dynamics directly with a set of
numerical experiments of MRI turbulence in
PP disks. We concentrate on the inner radii of these disks
($\sim 0.1-1$ au) so
that our simulation domain straddles both the bistable region and the inner
dead-zone boundary. Our aim is to understand the
interactions between real MRI turbulence and thermodynamics, and thus
test and extend previous exploratory work that modelled the former via a crude
diffusive process \citep{latter&balbus12}. Our 3D MHD simulations employ
 the Godunov code RAMSES \citep{teyssier02,fromangetal06}, in which
 the thermal energy equation has been accounted for and magnetic
 diffusivity appears as a function of temperature (according to an
 approximation of Saha's law). 
We focus on the optically thick disk midplane, 
and thus omit non-thermal sources of ionization. This also permits us
to model the disk in the cylindrical approximation. 

This initial numerical study is the first in a series, and thus
lays out our numerical tools, tests, and code-checks. We also verify
previous published results for fully turbulent disks and disks with
static dead zones \citep{nataliaetal10,lyra&maclow12}.
We explore the global radial profiles of thermodynamic variables
 in such disks, and the nature of the turbulent temperature fluctuations,
including a first estimate for the magnitude of the
turbulent thermal flux. We find that the dead zone can be effectively
heated by density waves generated at the dead zone boundary, and we
estimate the resulting wave damping and heating. Our simulations
indicate that the dead zone may be significantly hotter than most
global structure models indicate because of this effect. 
Another interesting result concerns the dynamics of the
dead zone interface itself. We verify that the interface is not static
and can migrate from smaller to larger radii. All such simulated MRI
`fronts' ultimately stall at a fixed radius set by the thermodynamic and
radiative profiles of the disk, in agreement with \citet{latter&balbus12}. These fronts move more quickly than predicted because they
propagate not via the slower 
MRI turbulent motions but by the faster density waves. 
Finally, we discuss instability near the dead zone
boundary.

The structure of the paper is as follows.
In \S{2} we describe the setup we used for the MHD simulations and give
our prescriptions for
the radiative cooling and thermal ionization.
We test our implementation of thermodynamical processes in models of
fully turbulent disks in \S{3}. There we also quantify the turbulent
temperature fluctuations and the turbulent heat flux.  
Section \S{4} presents results of resistive MHD simulations with dead
zones. Here we look at the cases of a static and dynamic dead zone separately. 
Rossby wave and other instabilities are briefly discussed in
\S{5}; we subsequently summarize our results and draw conclusions
in \S{6}.

\section{Setup}

We present in this paper a set of numerical simulations performed
using a version of the code RAMSES \citep{teyssier02,fromangetal06}
which solves the MHD equations on a 3D cylindrical and uniform grid. 
Since we are interested in the inner-edge behaviour at the mid-plane, we
discount vertical stratification and work under the
cylindrical approximation \citep{armitage98,
hawley01,steinacker&pap02}. In the rest of this section we present the
governing equations, prescriptions, main parameters, and initial and
boundary conditions. In order to keep the discussion as general
as possible, all variables and equations in this section are
dimensionless.
%% The setup closely resembles that used
%% by \citet{baruteauetal11}.
  
\subsection{Equations}
\label{fundeq}

Since we are interested in the interplay between dynamical and
thermodynamical effects at the dead zone inner edge, we solve the MHD
equations, with (molecular) Ohmic diffusion, alongside an energy equation.
%Being much smaller than Ohmic diffusion, 
We neglect the kinematic viscosity, that is much smaller than Ohmic diffusion, because of its minor role in the
MRI dynamics. However dissipation of kinetic energy is
fully captured by the numerical grid.
We adopt a cylindrical coordinate system $(R,\phi,Z)$ centred on
the central star:
\begin{eqnarray}
\frac{\partial \rho}{\partial t}+\del \bcdot (\rho \bb{v}) &=& 0 \\
\frac{\partial \rho \bb{v}}{\partial t}+\del \bcdot (\rho \bb{v} \bb{v} - \bb{B} \bb{B}) + \del P &=&
-\rho \del \Phi \\
\frac{\partial E}{\partial t}+\del \bcdot \left[ (E+P) \bb{v}
- \bb{B} (\bb{B} \bcdot \bb{v}) + {\cal F_{\eta}} \right] &=& -\rho\bb{v}\cdot\nabla \Phi
-{\cal L} \label{energy_eq} \\
\frac{\partial \vec{B}}{\partial t} - \bb{\nabla} \btimes
(\bb{v} \times \bb{B}) &=& - \bb{\nabla} \btimes (\eta \bb{\nabla}
\btimes \bb{B}) \label{induc}
\end{eqnarray}
where $\rho$ is the density, $\bb{v}$ is the velocity, $\bb{B}$ is the
magnetic field, and $P$ is the pressure. 
$\Phi$ is the gravitational potential.
In the cylindrical approximation, it is given by
$\Phi=-GM_{\star}/R$ where $G$ is the gravitational constant and
$M_{\star}$ is the stellar mass. In RAMSES, $E$ is the total energy,
i.e. the sum of kinetic, magnetic and internal energy $e_{th}$ (but it
does not include the gravitational energy). 
Since we use a perfect gas equation of state to close the former set
of equations, the latter is related to the pressure through the
relation $e_{th}=P/(\gamma-1)$ in which $\gamma=1.4$. The magnetic
diffusivity is denoted by $\eta$. As we only consider thermal
ionization, $\eta$ will depend on temperature $T$; its functional form
we discuss in the following section. Associated with that resistivity
is a resistive flux $\cal F_{\eta}$ that appears in the divergence of
Eq.~(\ref{energy_eq}). Its expression is given by Eq.~(23)
of \citet{1998RvMP...70....1B}. The $\mathcal{L}$ symbol denotes
radiative losses and it will be described in
section~\ref{thermo_sec}. Dissipative and radiative cooling
terms are computed using an explicit scheme. This method is valid if
the radiative cooling time scale is much longer than the typical
dynamical time (see section~\ref{init_sec}).
Finally, we have added a source term in the continuity equation that
maintains the initial radial density profile
$\rho_0$ \citep{NelsonGressel10,baruteauetal11}. It is such that 
\begin{equation}
\frac{\partial \rho}{\partial t}=-\frac{\rho-\rho_0}{\tau_{\rho}}.
\end{equation}
The restoring time scale $\tau_{\rho}$ is set to $10$ local orbits and
prevents the long term depletion of mass caused by the turbulent
transport through the inner radius. 
%Note that this procedure is performed at constant
%temperature. Thus, the internal energy is consistently modified. 
\subsection{Magnetic diffusion}
\label{saha}

In this paper, we conduct both ideal ($\eta=0$) and non-ideal
($\eta\neq 0$) simulations. In the latter case, we treat the disk as
partially ionized. In such a situation, the magnetic diffusivity is
known to be a function of the temperature $T$ and ionization fraction
$x_e$ \citep{blaes&balbus94}: 
\begin{equation}
\eta \propto  x_e^{-1} \, T^{1/2} \,.
\label{eta_eq}
\end{equation}
The ionisation fraction can be evaluated by considering the ionization
sources of the gas. 
In the inner parts of PP disks, the midplane electron fraction largely
results from thermal ionization (neglecting radioisotopes that are not
sufficient by themselves to instigate MRI). 
Non thermal contributions due to ionization by
cosmic \citep{cosmicSano00},
UVs \citep{PerezChiang11} and
X-rays \citep{1999ApJ...518..848I} 
are negligible because of the large optical thickness of the disk at 
these radii. In fact, the ionization is controlled by the low ionization
potential ($E_i \sim 1-5 eV$) alkali metals, like sodium
and potassium, the abundance of which is of order $10^{-7}$ smaller than
hydrogen \citep{pneuman&mitchell65,umebayashi&nakano88}. Then
$x_e$ can be evaluated using the Saha equation.  
When incorporated into Eq.~(\ref{eta_eq}), one obtains the following relation between
the resistivity and the temperature:
\begin{equation}
\eta \propto T^{-\frac{1}{4}} \exp{(T^{\star}/T)},
\label{eta_T}
\end{equation}
with $T^{\star}$ a constant. Because of the exponential, $\eta$ varies
very rapidly with $T$, and, in the context of MRI activation, can be
thought of as a `switch' around a threshold $T_\text{MRI}$. In
actual protoplanetary disks, it is well known that that temperature 
is around $10^3$~K \citep{balbus&hawley00,balbus11}. Taking
this into account, and for the sake of simplicity, we use in our
simulations the approximation:
\begin{equation}
\label{resist}
\eta(T)=
\begin{cases}
  \eta_0 \, \, \, \, \textrm{if} \, \, \, T<T_\text{MRI}\\
  0 \, \, \, \, \, \textrm{otherwise},
\end{cases} 
\end{equation}
where $T_\text{MRI}$ is the activation temperature for the onset of
MRI. Note that by taking a step function form for $\eta$ we must
neglect its derivative in Eq.~\eqref{induc}.

\subsection{Radiative cooling and turbulent heating}
\label{thermo_sec}
As we are working within a cylindrical model of a disk,
 without explicit surface layers, we model radiative losses
using the following cooling function: 
\begin{equation}
{\cal L} =\rho \sigma (T^4-T_\text{min}^4) \, .
\label{thermo_sec_eq}
\end{equation}
This expression combines a crude description of radiative
cooling of the disk ($-\rho \sigma T^4$) as well as irradiation by the
central star ($+\rho \sigma T_\text{min}^4$), in which $\sigma$ should be
thought of as a measure of the disk opacity (see also
section~\ref{init_sec}).
In the absence of any other heating or cooling sources, $T$
should tend toward $T_\text{min}$, the equilibrium
temperature of a passively irradiated disk. 
In principle, $T_\text{min}$ is a function of position and surface
density, but we take it to be uniform for simplicity. 
In practice, we found it has little influence on our results. 
Turbulence or waves ensure that the simulated temperature is always significantly
 larger than $T_\text{min}$. 
The parameter $\sigma$ determines the quantity of thermal energy the
gas is able to hold. 
For computational simplicity and to ease the physical interpretation
of our results, we take $\sigma$ to be uniform. In reality, $\sigma$
should vary significantly across the dust sublimation threshold $(\sim
1500$ K) (Bell \& Lin 1994), 
but an MRI front will always be much cooler $(\sim 1000$ K). As a consequence,
the strong variation of $\sigma$ will not play an important role in the dynamics.
Finally, we omit the radiative
diffusion of heat in the planar direction, again for simplicity though
in real disks it is an important (but complicated) ingredient. 

Our prescription for magnetic diffusion, Eq.~\eqref{resist}, takes care 
of the dissipation of magnetic 
energy when $T<T_\text{MRI}$. However, if the disk is sufficiently hot
the dissipation of magnetic energy (as with the kinetic energy) is not explicitly
calculated. In this case energy is dissipated on the grid;
our use of a total energy equation ensures that this energy is not
lost but instead fully transferred to thermal energy. It should be
noted that though this is perfectly adequate on long length and time
scales, the detailed flow structure on the dissipation scale will
deviate significantly from reality.

\subsection{Initial conditions and main parameters}
\label{init_sec}

Our 3D simulations are undertaken on a
uniformly spaced grid in cylindrical coordinates $(R,\phi,Z)$. 
The grid ranges over $R \in [R_0,8 \, R_0]$, $\phi \in [0,\pi/4]$ and
$Z \in[-0.3 \, R_0,0.3 \, R_0]$. Here the key length scale $R_0$ serves as
 the inner radius of our
disk domain and could be associated with a value between 0.2-0.4 au in
physical units. 
Each run has a resolution $[320,80,80]$, which has been shown in these
conditions to be sufficient for MHD turbulence to be sustained over
long timescales \citep{baruteauetal11}. For completeness, we
present a rapid resolution study in appendix~\ref{high_res_sim} that
further strengthens our results.

Throughout this paper, we denote by $X_0$ the value of the quantity
$X$ at the inner edge of the domain, i.e. at $R=R_0$. 
Units are chosen such that: 
\begin{equation*}
GM_{\star}=R_0=\Omega_0=\rho_0=T_0=1 \, ,
\end{equation*}
where $\Omega$ stands for the gas angular velocity at radius $R$. 
Thus all times are measured in inner orbits, so
a frequency of unity correspond to a period of one inner orbit.  
We will refer to the local orbital time at $R$ using the variable
$\tau_{orb}$. 

The initial magnetic field is a purely toroidal field whose profile is
built to exhibit no net flux and whose maximum strength corresponds to
$\beta=25$: 
\begin{equation}
B_\phi=\sqrt{\frac{2 P}{\beta}} \sin\left({\frac{2\pi}{Z_{max}-Z_{min}} Z}\right) \, .
\end{equation}
The simulations start with a disk initially in approximate
radial force balance (we neglect the small radial component of
the Lorentz force in deriving that initial state). Random velocities
are added to help trigger the MRI. Density and temperature profiles
are initialized with radial power laws: 
\begin{equation} 
\rho=\rho_0 {\left(\frac{R}{R_0}\right)}^{p} \quad \text{;} \quad
T=T_0 {\left(\frac{R}{R_0}\right)}^{q} \label{initT} \,,
\end{equation}
where $p$ and $q$ are free parameters. Pressure and temperature
are related by the ideal gas equation of state:
\begin{equation}
P=P_0 \left( \frac{\rho}{\rho_0} \right) \left( \frac{T}{T_0} \right) \, ,
\end{equation}
where $P_0$ depends on the model and is specified below.
We choose $p=-1.5$, which sets the initial radial profile for the density.
During the simulation, it is expected to evolve on the long secular
timescale associated with the large-scale accretion flow (except
possibly when smaller scale feature like pressure bump appear, see
section~\ref{non_ideal_mhd_sec}). 
Quite differently, the temperature profile evolves on the shorter
thermal timescale (which is itself longer than the dynamical timescale $\tau_{orb}$). 
It is set by the relaxation time to thermal equilibrium\footnote{One consequence is
that the ratio $H/R$ which measures the relative importance of thermal
and rotational kinetic energy, is no longer an input parameter, as in
locally isothermal simulations, but rather the outcome of a
simulation.}.

%% Following  \citeauthor{balbus&pap99} (\citeyear{balbus&pap99}, see their Eq.~(46))
%% the diffusive prescription account for the mean turbulent heating, replacing the large scale viscous stress
%% by the $R\phi$ component of the turbulent stress tensor:
As shown by \citet{1973A&A....24..337S}, turbulent heating in an
accretion disk is related to the $R\phi$ component of the turbulent stress
tensor $T_{R\phi}$:
\begin{equation}\label{Qplus}
Q_+=-T_{R\phi} \frac{\partial \Omega}{\partial \ln R} \sim 1.5 \Omega
T_{R\phi} \, .
\end{equation}
For MHD turbulence $T_{R\phi}$ amounts to \citep{balbus&pap99}:
\begin{equation}
T_{R\phi}=\mean{-B_R B_\phi + \rho v_R \delta v_\phi} \, .
\end{equation}
The $\mean{.}$ notation stands for an azimuthal, vertical and
time average, $\delta v_\phi$ is the flow's azimuthal deviation from
Keplerian rotation. The last scaling in equation \eqref{Qplus} assumes such
deviations are small, i.e. the disk is near
Keplerian rotation. Balancing that heating rate by the cooling
function $\mathcal{L} \sim \rho \sigma T^4$ gives a relation for the disk
temperature (here we neglect $T_{\rm min}$ in the cooling function):
\begin{equation}
\sigma \rho T^4 \simeq 1.5 \Omega T_{R\phi} \, .
\label{thermal_eq}
\end{equation}
Using the standard $\alpha$ prescription for illustrative purposes, we
write $T_{R\phi}=\alpha \mean{P}$ 
which defines the Shakura-Sunyaev $\alpha$ parameter, a constant in the
classical $\alpha$ disk theory (Shakura \& Sunyaev 1973). 
Since $\Omega \propto R^{-1.5}$, Eq.~(\ref{thermal_eq}) suggests that
$T \propto R^{-0.5}$ in equilibrium for uniform $\sigma$ values. 
We thus used $q=-0.5$ in our initialization of $T$.

We can use a similar reasoning to derive an estimate for the thermal
time-scale. Using the turbulent heating rate derived
by \citet{balbus&pap99} and the definition of $\alpha$, the mean
internal energy evolution equation can be approximated by
\begin{equation}
\label{eqtemp}
\frac{\partial e_{th}}{\partial t} \sim
\frac{\mean{P}}{(\gamma-1)\tau_{heat}} =1.5 \Omega \alpha \mean{P} -
     {\cal L} \, ,
\end{equation}
which yields immediately
\begin{equation}
\tau_{heat}^{-1} \sim 1.5 \Omega (\gamma -1) \alpha.
\end{equation}
For $\alpha=0.01$, the heating time scale is thus about $25$ local
orbits. The cooling time-scale $\tau_{cool}$ is more difficult to
estimate, but near equilibrium should be comparable to
$\tau_{heat}$. 

In order to obtain a constraint on the parameter $\sigma$ we use 
the relations $c_s^2=\gamma P/\rho \approx H^2\Omega^2$, where $c_s$
is the sound speed and $H$ is the disk scale height. 
At $R=R_0$ approximate thermal equilibrium, Eq.~(\ref{thermal_eq}),
can be used to express $\sigma$ as a function of disk parameters:
\begin{equation}
\sigma\approx \frac{3}{2} \frac{\alpha
R_0^2 \Omega_0^3}{\gamma\,T_0^4} \left( \frac{H_0}{R_0} \right)^2 \,.
\label{sigma_eq}
\end{equation}
Now setting $\alpha \sim 10^{-2}$ (and moving to numerical units)
ensures that $\sigma$ only depends on the ratio $H_0/R_0=c_0/(R_0 \Omega_0)$. We investigate two
cases: a `hot' disk with $H_0/R_0=0.1$ which yields
$\sigma=1.1\times10^{-4}$, and a `cold' disk with $H_0/R_0=0.05$ and
thus $\sigma=2.7\times 10^{-5}$. The two parameter choices for
$\sigma$  will be referred to as the $\sigma_{hot}$ and
$\sigma_{cold}$ cases in the following.
Using the relation between pressure and sound speed respectively yields $P_0=7.1 \times
10^{-3}$ and $P_0=1.8 \times 10^{-3}$ for
the two models. The surface density profile in the simulations
is given by a power law:
\begin{equation}
\Sigma=\rho_0 H_0\left(\frac{R}{R_0}\right)^{p+q/2+3/2} \, .
\end{equation}

%% balance heating vs cooling at the top
%% \begin{equation}
%% \frac{\sigma_b T^4}{\tau}=1.5 \alpha \Omega P H
%% \end{equation}
%% $T=1200K$ \Sigma=1700~g.cm^{-2} en R=1AU$
%% $\tau=130$ at $z=H$ dans le cas $\sigma_{hot}$.
%% \begin{equation}
%% \tau=\kappa \Sigma_{gas} d
%% \end{equation}
%% With $d$ is the dust to gas ratio.
%% $\kappa=130/17 \simeq 7.5$~cm$^{2}.g$^{-1}$$.

In resistive simulations, the parameters $T_\text{min}$,
$T_\text{MRI}$ and $\eta_0$ (when applicable) need to be specified for the
runs to be completely defined. 
We used  $T_\text{min}=0.05 \, T_0$, which means that the temperature in
the hot turbulent innermost disk radius is 20 times that of a cold passive
irradiated disk. 
As discussed in section~\ref{thermo_sec}, $T_\text{min}$ is so small
that it has little effect on our results. 
The values of $T_\text{MRI}$ and $\eta_0$ vary from model to model and
will be discussed in the appropriate sections.  

In order to assess how realistic our disk
model is, we convert some of its key
variables to physical units. This can be done as
follows. We consider a protoplanetary disk orbiting around a half
solar mass star, with surface density $\Sigma \sim 10^3$~g.cm$^{-2}$
and temperature $T \sim 1500$ K at $0.1$ AU (Given the radial
profile we choose for the surface density, this would correspond to a
disk mass of about seven times the minimum mass solar nebular within
100 AU of the central star). For this set of parameters, 
we can use Eq.~(\ref{sigma_eq}) to calculate $\sigma=1 \times 10^{-9}$
in cgs units for model $\sigma_{\textrm{cold}}$, using $\alpha \sim
10^{-2}$. In the simulations, the vertically integrated cooling rate
is thus given by the cooling rate per unit surface: 
\begin{equation}
Q_-=\Sigma \sigma T^4=5.3 \times 10^6 \textrm{
erg.cm}^{-2}.\textrm{s}^{-1} \, .
\label{qminus_simus_eq}
\end{equation}
This value can be compared to the cooling rate of a typical $\alpha$
disk with the same parameters \citep{chambers09}:
\begin{equation}
Q_-^{\textrm{PP}}=\frac{8}{3} \frac{\sigma_b}{\tau} T^4 \, ,
\end{equation}
where $\sigma_b$ is the Stefan-Boltzman constant and
$\tau=\kappa_0 \Sigma/2$ in which $\kappa_0=1$ cm$^2$.g$^{-1}$
stands for the opacity. Using these figures, we obtain
$Q_-^{\textrm{PP}}=1.5  \times 
10^6 \textrm{ erg.cm}^{-2}.\textrm{s}^{-1}$, 
i.e. a value that is close (given the level of approximation involved)
to that 
used in the simulations\footnote{ With our choice of parameters,
we have $Q_- \sim R^{-2.5}$ and $Q_-^{PP} \sim R^{-1.5}$. This means
that the agreement between $Q_-$ and $Q_-^{PP}$ improves with
radius. At small radial distances, dust sublimates and our model
breaks donwn (see section~\ref{conclusion_sec})} (see
Eq.~(\ref{qminus_simus_eq})). We caution 
that this acceptable agreement should not be mistaken as a proof that
we are correctly modelling all aspects of the disk's
radiative physics. The cooling function
we use is too simple to give anything more
than an idealized thermodynamical model. It only
demonstrates the consistency between
the thermal timescale we introduce in
the simulations and the expected cooling timescale in proto-planetary
disks.

\subsection{Buffers and boundaries}
\label{buffers_sec}

Boundary conditions are periodic in $Z$ and $\phi$ while special care
has been paid to the radial boundaries. Here $v_R$, $v_z$, $B_{\phi}$,
$B_z$ are set to zero, $B_R$ is computed to enforce magnetic flux
conservation, $v_{\phi}$ is set to the Keplerian value, and finally
temperature and density are fixed to their initial values. 
 
As is common in simulations of this kind, we create two buffer zones
adjacent to the inner and outer boundaries in which
 the velocities are damped toward their
boundary values in order to avoid sharp discontinuities. 
The buffer zones extend from $R=1$ to $R=1.5$ for the
inner buffer and from $R=7.5$ to $R=8$ for the outer one. 
For the same reason, a large resistivity is used in those  buffer
zones in order to prevent the magnetic field from accumulating next to
the boundary. As a result of this entire procedure, turbulent activity
decreases as one approaches the buffer zones. This would occasionally
mean the complete absence of turbulence in
the region close to the inner edge because of inadequate resolution.  
To avoid that problem, the cooling parameter $\sigma$ gradually increases
with radius in the region $R_0<R<2R_0$. In
appendix \ref{buff_appendix}, we give for completeness the functional
form of $\sigma$ we used. These
parts of the domain that corresponds to the inner and outer buffers
are hatched on all plots of this paper.

\section{MHD simulations of fully turbulent PP disks}
\label{ideal_mhd_sec}

We start by describing the results we obtained in the `ideal' MHD limit:
$\eta=0$ throughout the disk. As a consequence, there is no dead zone
and the disk becomes fully turbulent as a result of the MRI. The purpose
of this section is (a) to describe the thermal structure of
the quasi-steady state 
that is obtained, (b) to check the predictions of simple alpha models,
and (c) to examine the small-scale and short-time thermodynamic
fluctuations of the gas, especially with respect to their role in
turbulent heat diffusion.
Once these issues are understood we can turn with confidence to
the more complex models that exhibit dead-zones.

\subsection{Long-time temperature profiles}
\label{tests}

\begin{figure}
\begin{center}
\includegraphics[width=7cm]{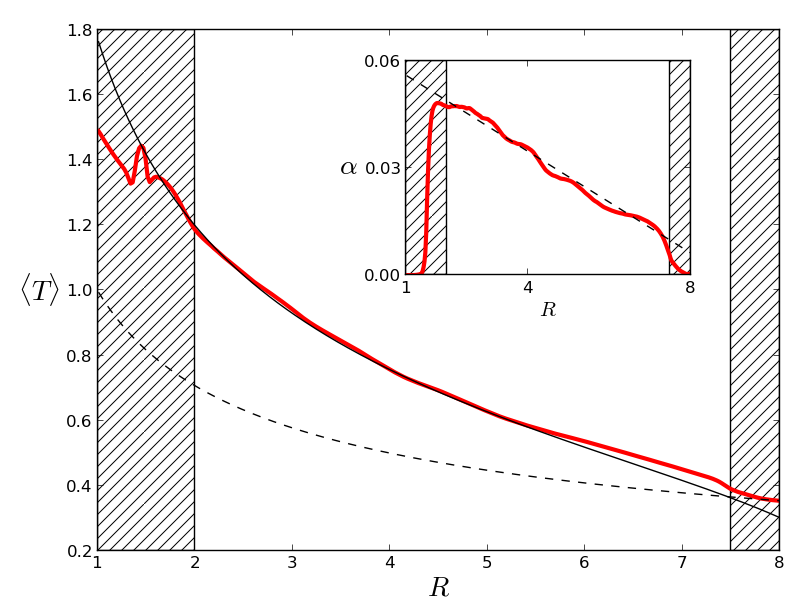}
\includegraphics[width=7cm]{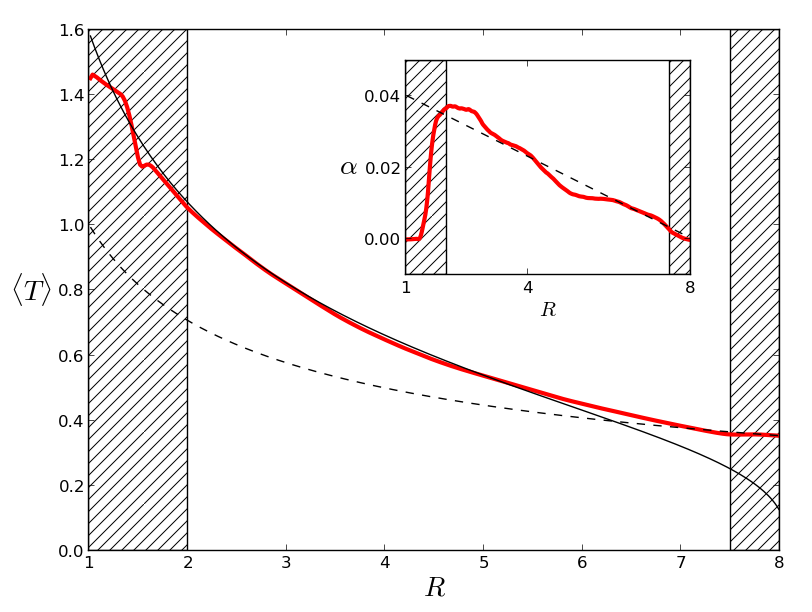}
\caption{Temperature profiles $\langle T \rangle$ averaged  over $900$
  inner orbits (red curves). Black plain lines show their corresponding theoretical
  profiles. Black dashed lines show their corresponding initial
  profiles. Top panel: $\sigma_{cold}$ case. Bottom panel:
  $\sigma_{hot}$ case.
  The subframes inserted in the upper right of each panel show the alpha
  profiles obtained in both cases (red lines).
  The dashed lines show the analytic profiles
  used to compute the theoretical temperature profile (see text), given by 
  $\alpha=6.3 \times 10^{-2}-7.1   \times 10^{-3} R/R_0$ and $\alpha=4.6 
  \times 10^{-2}-5.7 \times 10^{-3} R/R_0$ respectively.}  
\label{verifT}
\end{center}
\end{figure}

We discuss here the results of the
 two simulations that correspond to $\sigma_{hot}$ and
$\sigma_{cold}$.
We evolve the simulations not only for a long enough time 
for the turbulence transport
properties to reach a quasi-steady state but also for the
thermodynamic properties to have also relaxed. Thus the simulations are
evolved for a time much longer than the thermal time of the gas, $\tau_{heat}
\approx 25$ local orbits. Here, we average over nearly 1000 inner orbits (about
$2\tau_{heat}$ at the outermost radius $R=7$). 
 
In figure 1 we present the computed radial temperature profiles and the radial profiles of $\alpha$ for
the two simulations. Note that $\alpha$ is clearly not constant in space. In both
models, $\alpha$ is of order a few times $10^{-2}$ and decreases
outward. We caution here that this number and radial profiles are to
be taken with care. The $\alpha$ value is well known to be affected by
numerical convergence \citep{beckwithetal11,sorathiaetal12} as well as
physical convergence issues
\citep{fromangetal07,lesur&longaretti07,simon&hawley09}. This may  
also influence the temperature because it depends on the turbulent activity.

The disk temperature $T$ rapidly departs from the initialized power
law given by Eq.~(\ref{initT}). In both plots the averaged simulated
 temperature decreases faster than $R^{-0.5}$. This is partly a
result of $\alpha$ decreasing with radius, and a consequent reduced
 heating with radius. The disk aspect ratios are
close to the targeted values: $H/R \sim 0.15$ and $0.08$ in model $\sigma_{hot}$
and $\sigma_{cold}$ respectively, slightly increasing with radius in
both cases.

One of our goals is
to check the implementation of the thermodynamics, and to ensure that
we have reached thermal equilibrium. To accomplish this 
we compare the simulation temperature profiles 
 with
theoretical temperature profiles computed according to the results of
\citet{balbus&pap99}. The theoretical temperature profile
can be deduced from thermal equilibrium, Eq.~(\ref{thermal_eq}), in which we
now include the full expression for the cooling function
(i.e. including $T_{\rm min}$):
\begin{equation}
T= \left( T_\text{min}^4 + \frac{3}{2} \frac{\alpha \Omega \mean{P}}{\sigma
  \mean{\rho}} \right)^{1/4} \, . 
\label{temp_th_eq} 
\end{equation}
Using the simulated $\alpha$ profiles as inputs, we could then
calculate radial profiles for $T$. In fact,
we used linear fits of $\alpha$ (shown as a dashed
line) in Eq.~(\ref{temp_th_eq}), for simplicity.
The theoretical curves are compared with the
simulation results in figure~\ref{verifT}.
The overall good agreement validates our implementation of the source
term in the energy equation and also demonstrates that we accurately
capture the turbulent heating.
 It is also a numerical confirmation that turbulent
energy is locally dissipated into heat in MRI--driven turbulence
 and that our
simulations have been run sufficiently long to achieve thermal equilibrium.  
During the steady state phase of the simulations, the mass
loss at the radial boundaries is very small: the restoring rate
is  $\dot{\rho} \sim 5 \times 10^{-4} \rho_0 \Omega_0$ on average
on the domain.

\subsection{Turbulent fluctuations of temperature}

\begin{figure}
\begin{center}
\includegraphics[width=7cm]{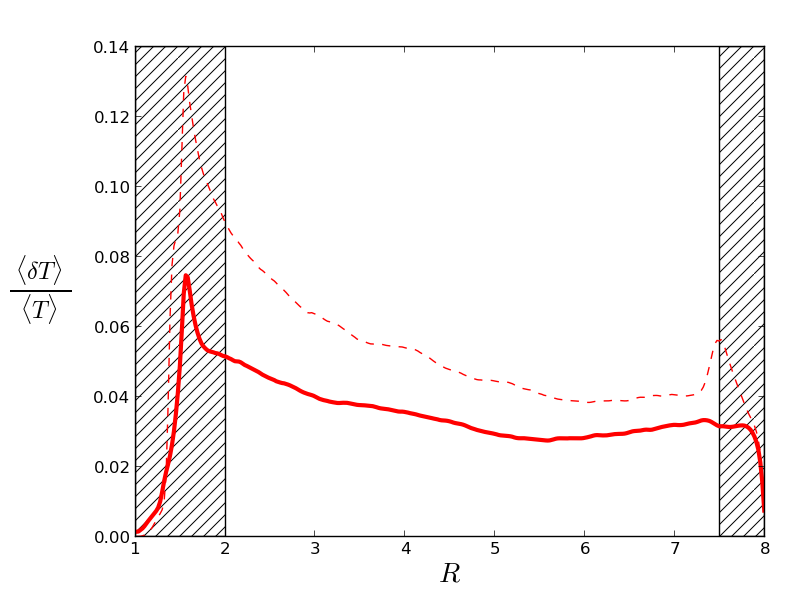}
\caption{Temperature standard deviation profiles averaged over $900$
  inner orbits. The plain line show the result from the $\sigma_{hot}$
  case and the dashed curve show the result from the $\sigma_{cold}$
  run.} 
\label{dtdtdt}
\end{center}
\end{figure}

\begin{figure}
\begin{center}
\includegraphics[width=7cm]{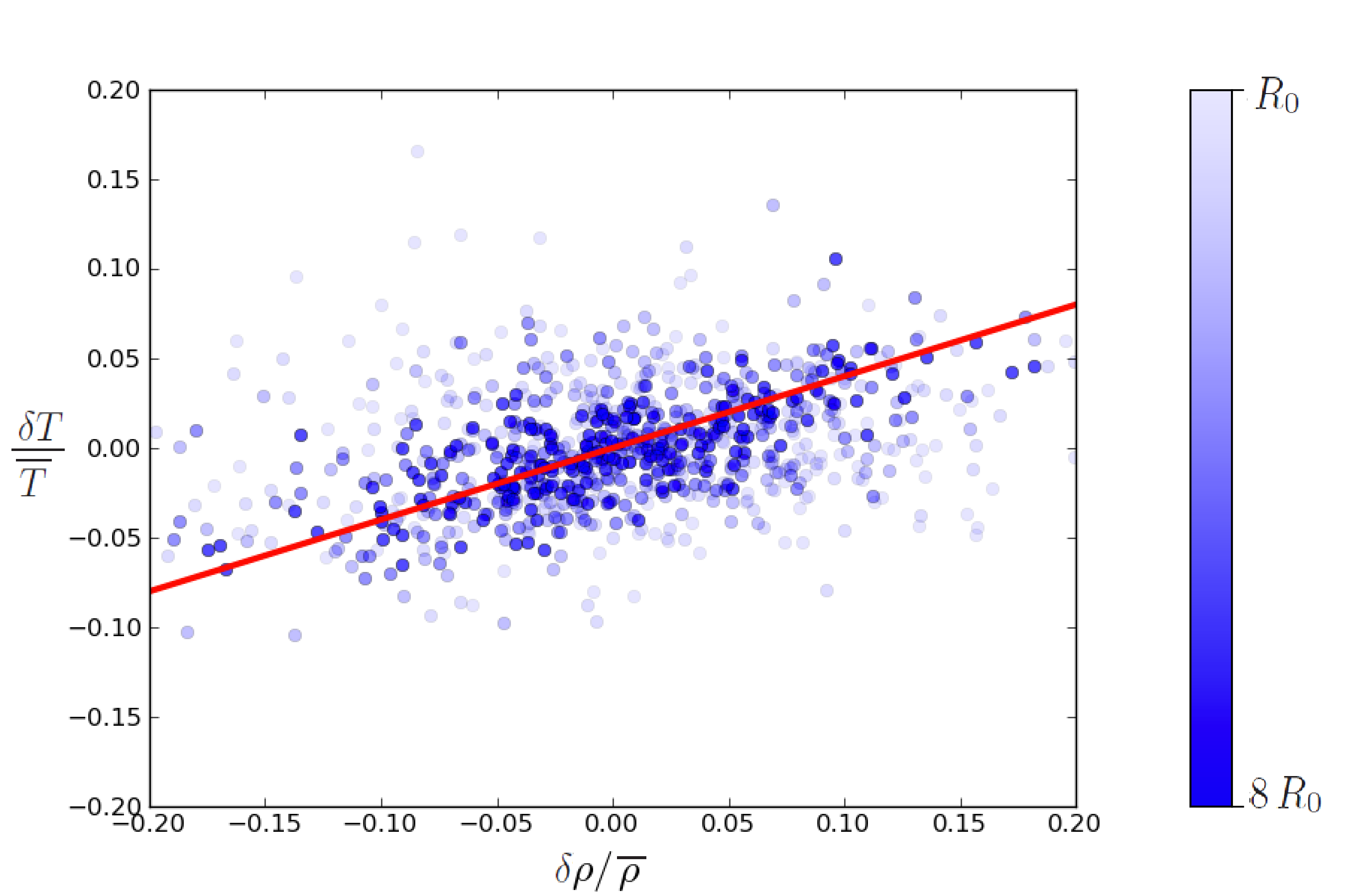}
\caption{Blue dots map the scaling relation between temperature and
  density fluctuations $\delta T$ and $\delta \rho$ in the
  $\sigma_{hot}$ simulation. $900$ events are randomly selected in the
  whole domain (except buffer zones) among $900$ inner orbits. The red
  line shows the linear function (slope $=\gamma-1)$) corresponding to the adiabatic scaling.}  
\label{adia}
\end{center}
\end{figure}

We now focus on the local and short time evolution of $T$,
by investigating the fluctuations of density and temperature
once the mean profile has reached a quasi steady state.
We define the temperature fluctuations by $\delta
T=T(R,\phi,Z,t)-\overline{T}(R,t)$, where an overline indicates an
azimuthal and vertical average. The magnitude of the fluctuations are
plotted in figure~\ref{dtdtdt}. 
In the $\sigma_{cold}$ case, the fluctuations range between $4$ and $8$\%
while they vary between $3$ and $5$\% in the $\sigma_{hot}$
case. The smaller temperature fluctuations in the latter case possibly
reflect the weaker turbulent transport in the latter case, while also
its greater heat capacity.
The tendency of the relative temperature fluctuations to decrease with
radius is due to $\alpha$ decreasing outward.

These temperature fluctuations can be due to different kinds of events
that act, simultaneously or not, to suddenly heat or cool the gas. Such
events can, for example, be associated with adiabatic compression or
magnetic reconnection. To disentangle these different
possibilities, we note that turbulent compressions, being primarily adiabatic,
 satisfy the following relationship between $\delta T$ and
the density fluctuations $\delta \rho$:
\begin{equation}
\frac{\delta T}{\overline{T}}=(\gamma-1) \frac{\delta \rho}{\overline{\rho}}.
\end{equation}
On figure~\ref{adia} we plot the distribution of a series of
fluctuation events randomly selected during the $\sigma_{hot}$ model
in the $(\delta \rho/\overline{\rho},\delta T/\overline{T})$ plane. 
The dot opacity is a function of the event's radius: darker points
stand for outer parts of the disk. The events are scattered around the
adiabatic slope (represented with a red line), but display a large
dispersion that decreases as radius increases. The difference between
the disk cooling time and its orbital period is so large that the
correlation should be much better if all the heating/cooling events
were due to adiabatic compression. This implies that temperature
fluctuations are mainly the result of isolated
 heating events such as magnetic
reconnection. This suggestion is supported by
the larger dispersion at short distance from the star, where turbulent
activity is largest. However, a detailed study of the heating
mechanisms induced by MRI turbulence is beyond the scope of this
paper. Indeed, the low resolution and the uncontrolled dissipation at
small scales -- important for magnetic reconnection --
 both render any detailed analysis of the problem difficult. 
Future local simulations at high resolution and with explicit
dissipation will be
performed to investigate that question further.

\begin{figure}
\begin{center}
\includegraphics[width=7cm]{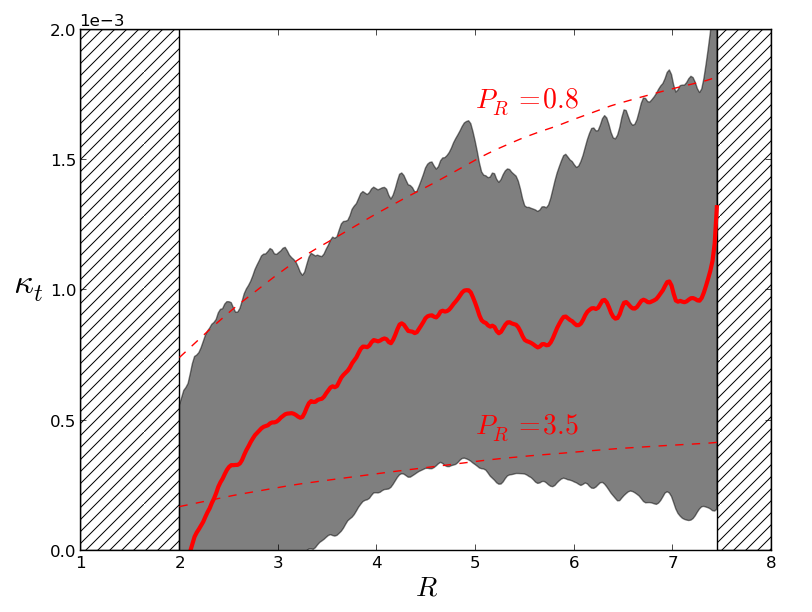}
\caption{Thermal diffusivity's radial profile averaged over $900$ inner
  orbits. Dashed lines show the mean thermal diffusivity for two
  homogeneous $\alpha_T$ ($\alpha_T=0.02$ and $\alpha_T=0.004$). The
  grey area delimit one standard deviation around the mean thermal
  diffusivity.}  
\label{kappa}
\end{center}
\end{figure}

As shown in \citet{balbus04}, MHD turbulence may also induce thermal
energy transport through correlations between the temperature
fluctuations $\delta T$ and velocity fluctuations $\vec{\delta
  v}$. For example, the radial flux of thermal energy is given by Eq.~(5) of \citet{balbus04}:
\begin{equation} 
F_{turb}=\frac{\overline{\rho} c_0^2}{\gamma-1}  \overline{\delta T
  \delta v_R } \, .
\label{fluxeq}
\end{equation}
By analogy with molecular thermal diffusivity, we quantify the
turbulent efficiency to diffuse heat using a parameter $\kappa_T$: 
\begin{equation}
\mean{F_{turb}}=- \frac{\kappa_T \mean{\rho} c_0^2}{\gamma (\gamma-1)} \frac{\partial \langle T
  \rangle}{\partial R} \, .
\label{kappaeq}
\end{equation}
To make connection with standard $\alpha$ disk models, we introduce
the parameter $\alpha_T$ which is a dimensionless measure of $\kappa_T$: 
\begin{equation}
\label{alphaT}
\alpha_{T}=\frac{\kappa_T}{ \langle c_s \rangle \langle H \rangle} \, ,
\end{equation}
as well as the turbulent Prandtl number $P_R=\alpha/\alpha_T$ that
compares turbulent thermal and angular momentum transports. We measure
the turbulent thermal diffusivity in the $\sigma_{hot}$ model and plot
in figure~\ref{kappa}
its radial profile and statistical deviation. 
For comparison, we also plot two radial profiles 
of $\kappa_T$ that would result from constant $\alpha_T$, chosen such
that they bracket the statistical deviations of the simulations
results. They respectively correspond to $P_R \sim 0.8$ and $P_R \sim 3.5$
(assuming a constant $\alpha=0.02$). The large deviation around the
mean value prevents any definitive and accurate measurement of $P_R$,
but our data suggest that $P_R$ is around unity. This is relatively
close to $P_R=0.3$ that \citet{pierensetal12} used to investigate,
with a diffusive model, how the turbulent diffusivity impacts on planet
migration. 

Turbulent diffusion must be compared to the radiative
  diffusion of temperature.
In the optically thick approximation, the radiative
  diffusivity is $\propto T^3$.
One can establish, using Eq.~(\ref{thermal_eq}), that the
  dimensionless measure of radiative diffusivity $\alpha_{rad}$,
  defined similarly to $\alpha_T$ (see Eq.~(\ref{kappaeq}) and
  Eq.~(\ref{alphaT})), is equal to a few $\alpha$. 
The turbulent transport of energy is then comparable to
  radiative transport in PP disks (see also Appendix A in Latter \&
  Balbus 2012).
%% \textbf{The radiative diffusion time over the domain vertical size is then $\Delta t_{rad} = 0.6^2 / \kappa_t \sim 70$ inner orbits.
%% It is shorter than the local cooling time for $R>2$, as neglecting the radiative diffusion (see Eq~(\ref{thermo_sec_eq})) requires.}
Nevertheless, radiative transport is neglected in our simulations. As
indicated by the discussion above, this shortcoming should be
rectified in future. Note that radiative MHD simulations are
challenging and \citet{2013A&A...560A..43F}
performed the first global radiative MHD simulations of
protoplanetary disks only very recently. Previous studies had been confined to the
shearing box approximation
\citep{2006ApJ...640..901H,2010MNRAS.409.1297F,HiroseTurner11} because
of the inherent numerical difficulties of radiative simulations.

%As we show
%in the Appendix {\bf (TBD)}, the radiative flux of thermal energy can
%be expressed as
%\begin{equation}
%F_R=\frac{\sigma T^3}{\mean{\rho}} \frac{\partial \mean{T}}{\partial
%  R} =  \frac{\kappa_R \mean{\rho} c_0^2}{\gamma (\gamma-1)} \frac{\partial \langle T
%  \rangle}{\partial R} \, ,
%\end{equation}
%where the second equality serves as a definition for the radiative
%diffusion coefficient $\kappa_R$. We can expressed the latter using a
%dimensionless coefficient $\alpha_R$ through $\kappa_R=\alpha_R c_s
%H$. Using Eq.~(\ref{thermal_eq}) as well as the definition of
%$\alpha$, we find

%% \textbf{(** the original argument is problematic! Where is the lengthscale
%%   coming out of the dTdR? Can't just replace the dTdR with
%%   T. **)}
 
%\begin{equation}
%\alpha_R=\frac{3}{2} (\gamma -1) \alpha \, .
%\end{equation}
%As shown above, turbulent transport of thermal energy is such that
%$\alpha \sim \alpha_T$. Thus, we obtain
%\begin{equation}
%\frac{\alpha_R}{\alpha_T} \sim \frac{3}{2} (\gamma -1) \sim 1 \, .
%\end{equation}
%This means that turbulent transport of energy can not be neglected and
%is always as important as radiative transport of energy in those parts
%of the disk where the temperature is set by turbulent dissipation.

%%% Local Variables: 
%%% mode: latex
%%% TeX-master: "main"
%%% End: 

\section{MHD simulations of PP disks with a dead zone}
\label{non_ideal_mhd_sec}

We now turn to the non-ideal MHD models in which the disk is composed of an
inner turbulent region and a dead zone at larger radii. To make
contact with previous work \citep{nataliaetal10,lyra&maclow12}, we
first consider the case of a resistivity that is only a function of
position. Such a simplification is helpful to understand
the complex thermodynamics of the dead zone before moving to the
more realistic case in which the resistivity is self-consistently
calculated as a function of the temperature using
Eq.~(\ref{resist}).

\subsection{The case of a static interface}
\label{static_interface_sec}

\begin{figure*}[!ht]
\begin{center}
\includegraphics[scale=0.4]{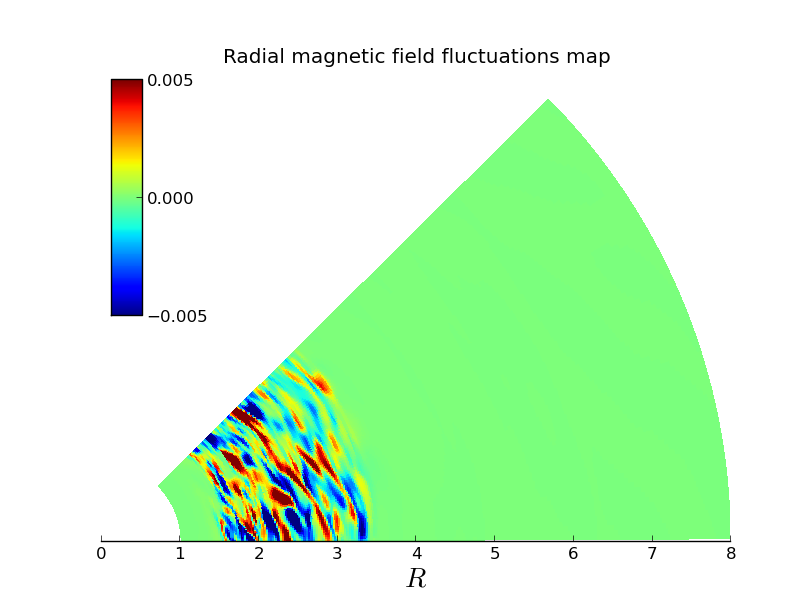}
\includegraphics[scale=0.4]{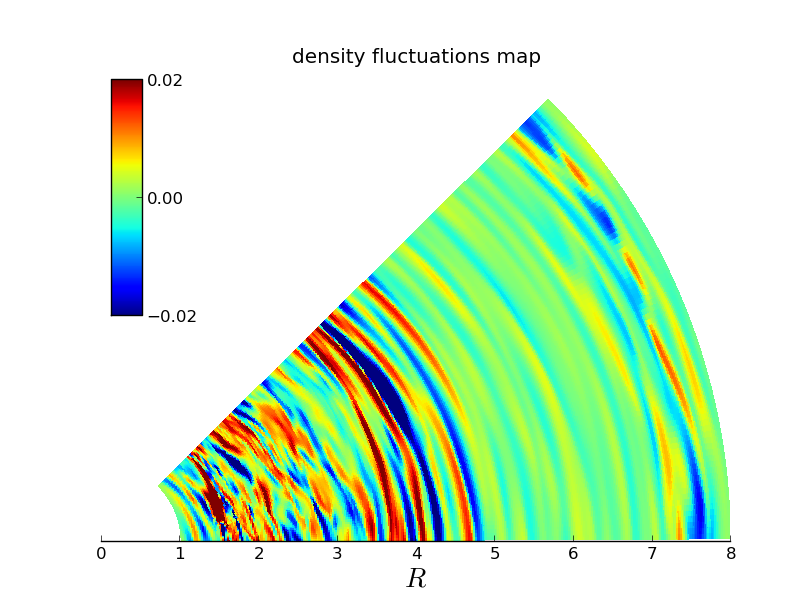}
\caption{$z=0$ planar maps of radial magnetic fields fluctuation (on the
  left) and density fluctuations (right plot) from the $\sigma_{hot}$
  simulation at $t=900+\tau_{cool}(R=7 R_0)$.} 
\label{fluctmapDZ}
\end{center}
\end{figure*}

\begin{figure}
\begin{center}
\includegraphics[scale=0.45]{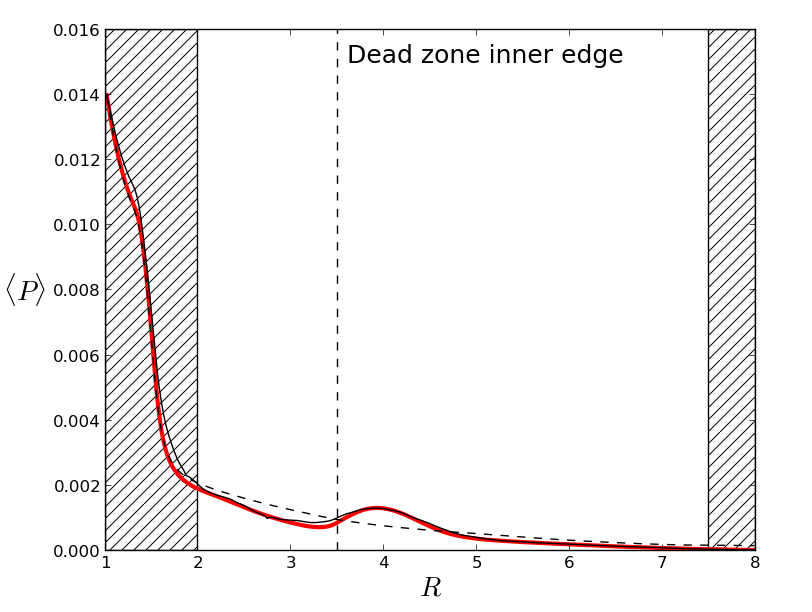}
\caption{Pressure profile averaged over $200$ orbits after
  $t=900+\tau_{cool}(R=7)$ in $\sigma_{Hot}$ case. The black solid
  line show the pressure profile $200$ inner orbits after freezing the
  temperature in the domain. The black dashed line plots the radial
  temperature profile in the ideal case (i.e. without a dead zone).}
\label{bump1}
\end{center}
\end{figure}

\begin{figure}
\begin{center}
\includegraphics[scale=0.45]{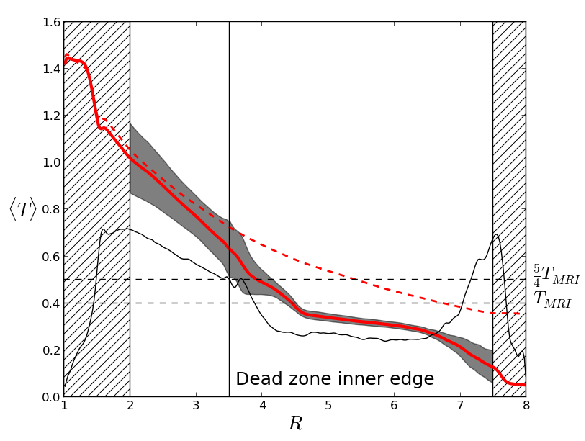}
\caption{The radial profile of temperature (solid red line) averaged over
  $200$ orbits after $t=900+\tau_{cool}(R=7 R_0)$. The red dashed line describes the 
  temperature profile from the ideal case and the wave-equilibrium profile
  is shown by the plain black line. The black dashed lines show the
  threshold value $T_\text{MRI}$ used in section~\ref{MRIfront} and
  $(5/4) T_{MRI}$. The grey area shows the deviation of
  temperature at $3$ sigmas. The vertical line shows the active/dead
  zone interface.}  
\label{tempDZ}
\end{center}
\end{figure}

At $t=900$, model $\sigma_{hot}$ has reached thermal equilibrium. We then
set $\eta=0$ for $R< 3.5$, and  $\eta=10^{-3}$ for $R \ge 3.5$, and
restart the simulation. The higher value of $\eta$ corresponds 
to a magnetic Reynolds number $R_m \sim c_s H/\eta \sim 10$. It is sufficient
to stabilize the MRI, and as expected, the flow becomes laminar
outward of $R=3.5$ in a few orbits. The
structure of the flow after $460$ inner orbits (which roughly amounts
to one cooling time at $R=7$) is shown in figure~\ref{fluctmapDZ}. The
left panel shows a snapshot of the radial magnetic field. The
turbulent active region displays large fluctuations which can easily
be identified. The interface between the active and the dead zone,
defined as the location where the Maxwell stress (or, equivalently,
the magnetic field fluctuations) drops to zero, stands at $R=3.5$
until the end of this \textit{static} non-ideal simulation. 

As soon as the turbulence vanishes, the sharp gradient of
  azimuthal stress at the interface drives a strong radial outflow. 
As identified by \citet{nataliaetal10} and
\citet{lyra&maclow12}, we observe the formation of a pressure and a
density maximum near the active/dead interface.
In figure~\ref{bump1}, we present the mean pressure perturbation we
obtained in that simulation.
The growth time scale of that structure is so short that the
source term in the continuity equation (see section~\ref{fundeq}) has no qualitative
impact because it acts on the longer time scale $\tau_{\rho}$. This feature persists even 
after we freeze the temperature and use a locally isothermal equation
of state as shown by the black solid line. We conclude that pressure
maxima are robust features that form at the dead zone inner edge
independently of the disk thermodynamics. 

We next turn to the temperature's radial profile. Because we expect the
dead zone to be laminar, we might assume the temperature to drop to a
value near
$T_\text{min}=0.05$ (see
section~\ref{thermo_sec}). However, the time averaged temperature
radial profile plotted in figure~\ref{tempDZ} shows this is far from
being the case. The temperature in the bulk of the dead
zone levels off at $T \sim 0.3$, significantly above
$T_\text{min}$. This increased temperature cannot be attributed to
Ohmic heating, since the magnetic energy is extremely small outward of
$R=3.5$ (at $R=4$, the magnetic energy has dropped by a factor of $10^2$
  compared to $R=3.5$, while at $R=5$ it
  has been reduced by about $10^3$). In contrast, there are significant hydrodynamic
fluctuations in the dead zone, as shown on the right panel of
figure~\ref{fluctmapDZ}. For example, density fluctuations
typically amount to $\delta \rho/\rho$ of the order of a few percent
at $R=5$. The spiral shape of 
the perturbation indicate that these perturbations could be
  density waves propagating in the dead zone, likely excited at the
  dead/active interface.
%% \textbf{We present a detailed analysis of their propagation in appendix~\ref{wave_heating_ap}.}
%% \textbf{It shows especially that waves are} excited \textbf{around} the dead/active interface and these propagate outward in
%% \textbf{to exclude the possibility that waves would be generated by
%% the outer boundary conditions, we have run an additional simulation with
%% identical parameters but with a much greater outer boundary, $R=22$ 
%% (thus the outer buffer zone extends from $R=21.5$).
%% The figure~\ref{damp} in Appendix~\ref{large_radius_app}
%% nontheless but also gives strong indications that waves are
%% probably generated at(or close to) the interface.
%% Further details of this calculation are given in Appendix~\ref{large_radius_app}}
Such a mechanism would be similar to the excitation of sound waves
 by the active surface layers observed in
vertically stratified shearing boxes simulations
\citep{fleming&stone03,oishi&maclow09} and also the waves seen in global
simulations such as ours \citep{lyra&maclow12} but performed using a
locally isothermal equation of state. However, as opposed to 
\citet{lyra&maclow12}, the Reynolds stress associated with this waves
is smaller in the dead zone than in the active zone, and amounts to
$\alpha \sim 10^{-4}$. Such a difference is likely due to the smaller
azimuthal extent we use here, as it prevents the appearance of a
vortex \citep[see section~\ref{rossby_wave_sec}
  and][]{lyra&maclow12}.

\subsection{Wave dissipation in the dead-zone}

It is tempting to use 
Eq.~(\ref{temp_th_eq}) to estimate the temperature that would result
if the wave fluctuations were \emph{locally} dissipated into heat (as is the
case in the active zone). However, we obtain $T \sim 0.1$ which
significantly underestimates the actual temperature.
 This is probably because the density waves we
observe in the dead zone do not dissipate locally. 
We now explore an alternative
model that describes their effect on the thermodynamic structure of
the dead zone.

We assume that waves propagate
adiabatically before dissipating in the form of weak
shocks in the dead zone; this assumption is consistent with measured
Mach
numbers ($\sim 0.1$) in the bulk of our simulated dead zone.
A simple model for the weak shocks, neglecting dispersion,
 yields a wave heating rate $Q_w^+$:
\begin{equation}
Q_w^+=\frac{\gamma(\gamma + 1)}{12} \langle P \rangle  
  \left( \frac{\delta \rho}{\overline{\rho}} \right)^3  f \, , 
\label{shoki}
\end{equation}
where $f$ stands for the wave frequency
\citep{ulmschneider70,camille}. 
The derivation of that
expression is presented in appendix~\ref{wave_heating_ap}. Balancing
that heating by the local cooling rate $\langle \rho \rangle \sigma
T^4$ yields an estimate of the temperature's radial profile. 
\begin{equation}
\sigma T^4 =
  \frac{\gamma + 1}{12} \left( \frac{H}{R} \right)^2 (R \Omega)^2  
  \left( \frac{\delta \rho}{\overline{\rho}} \right)^3  f \, ,
\label{temp_dz_eq}
\end{equation}
where the relation between the pressure and the disk scale height has
been used. This expression provides an estimate of the dead-zone
temperature. If one assumes that the waves are excited around the
dead-zone interface, the frequency should be of order the
inverse of the correlation time $\tau_{c}$ of the turbulent
fluctuations at that location. Such a short period for the waves, much
shorter than the disk cooling time, is consistent with the hypothesis
that these waves behave adiabatically. Local simulations of \citet{fromang&pap06} suggest
$\tau_{c}/T_{orb} \sim 0.15$. Given the units of the simulations, this
means that we should use $f \sim 1$. Taking $\delta \rho/\rho \sim
0.1$ and $H/R \sim 0.1$, this gives $T \sim 0.25$, at $R=5$, which
agrees relatively well with the measured value.
The agreement certainly
 supports the assumption that waves are generated at the interface.

In order to make the comparison more precise, we plot in figure~\ref{tempDZ} the
radial profile of the temperature, computed using
Eq.~(\ref{temp_dz_eq})\footnote{We used the azimuthally and temporally
averaged simulation data for $H/R$ and $\delta \rho/\rho$ when
computing the temperature using Eq.~(\ref{temp_dz_eq})}.
The agreement with the temperature we measure in the simulation is
more than acceptable in the bulk of the dead zone. There are however
significant deviations at $R \ge 7$, i.e. next to the outer
boundary. Associated with these deviations, we also measured large,
but non wave--like values of the quantity $\delta \rho/\rho$ at that
location (see figure~\ref{fluctmapDZ}). 

In order to confirm that the
outer-buffer zone is responsible for these artefacts, we ran
an additional simulation with identical parameters but with a much
wider radial extent. The outer-radial boundary is located at
$R=22$, and thus the outer-buffer zone extends from 
$R=21.5$ to $R=22$. The additional simulation shows that
indeed the suspicious temperature bump moves to
the outer boundary again. 
The temperature in the bulk of the dead zone, on the other hand, remains
unchanged and is in good agreement with the model
$\sigma_{hot}$. Further details of this calculation are given in
Appendix~\ref{large_radius_app}. 

In Appendix C we show that the density
wave theory can also account for the density
fluctuations as a function of radius. As waves
propagate, their amplitudes decrease, not least because their energy content is
converted into heat. Overall, the good match we have obtained 
between the wave properties' radial profiles (both for the temperature
and density fluctuations) and the analytical prediction strongly favour
our interpretation that density waves are emitted at the active/dead
zone interface and rule out any numerical artefacts that might be
associated with the domain outer boundary.

\begin{figure}
\begin{center}
\includegraphics[width=7cm]{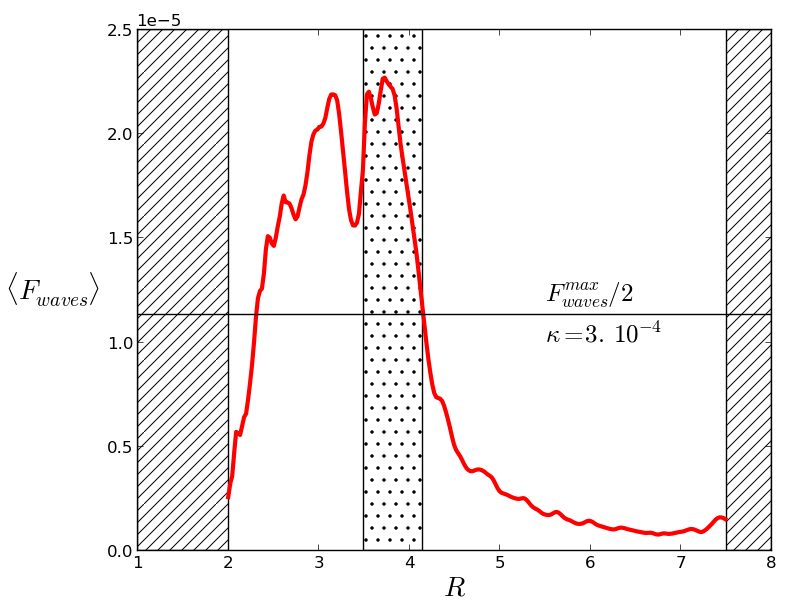}
\caption{Turbulent thermal flux profile averaged over $200$ inner
  orbits after $t=900+\tau_{cool}(R=7)$ from the $\sigma_{hot}$
  run. The central dotted region locates the percolation region
  behind active/dead zone interface.}  
\label{divDZ}
\end{center}
\end{figure}

The effect of the waves can be viewed as a flux of thermal energy
carried outward. Such a thermal flux can be quantified using 
Eq.~(\ref{fluxeq}) and its radial profile in model $\sigma_{hot}$ is shown
in figure~\ref{divDZ}. It is positive and decreasing in the dead zone
which confirms that thermal energy is transported from the active zone
and deposited at larger radii.
While it is rapidly decreasing with $R$,
the important point of figure~\ref{divDZ} is that it remains important
in a \textit{percolation region} outward of the dead zone inner edge
(located at $R=3.5$).
Most of the wave energy that escapes the
active region is transmitted to the dead zone over a \textit{percolation length},
the size of the percolation region.
We define the \textit{percolation region} as the region inside the dead zone where
the thermal transport amounts to more that half its value in the active zone.
As shown by figure~\ref{divDZ}, the percolation length $l \sim
0.5$ in model $\sigma_{hot}$, which translates to about one scale
height at that location.

Though the heat flux is greatest within about $H$ of the dead zone
interface, the action of the density waves throughout our simulated dead zone keeps
temperatures significantly higher than what would be the case in 
radiative equilibrium. This would indicate that much of the dead zone
in real disks would be hotter than predicted by global structure
models that omit this heating source. In particular, this should have
important implications for the location of the ice line, amongst other
important disk features. 

%% \textbf{(** reader may wonder, why invoke a complicated shock-wave
%%   theory? Why not just measure hydrodynamic alpha in the dead zone, as
%%   Lyra does, and plug that into the heat balance? I think we should do
%% both, actually, to show consistency. Unless, of course, the alpha from
%% density waves doesn't work - and there are reasons for why it might
%% not. The spiral waves may not dissipate locally... **)}

%To conclude this section, we report that density waves
%generated at the active/dead zone interface can transport significant
%heat into the dead zone. A simple theory can be developed to account
%for that remnant heating and can be validated by the results of the
%simulations. The extended $\sigma_{hot}$ simulation suggests that the 
%heated region extends quite far away from the interface: the temperature in
%the dead zone still exceeds twice the value of $T_{min}$ at $R=12$,
%i.e. about $20$ scale heights away from the interface.

\subsection{The case of a dynamic interface}
\label{MRIfront}

\begin{figure*}[!ht]
\begin{center}
\includegraphics[scale=0.28]{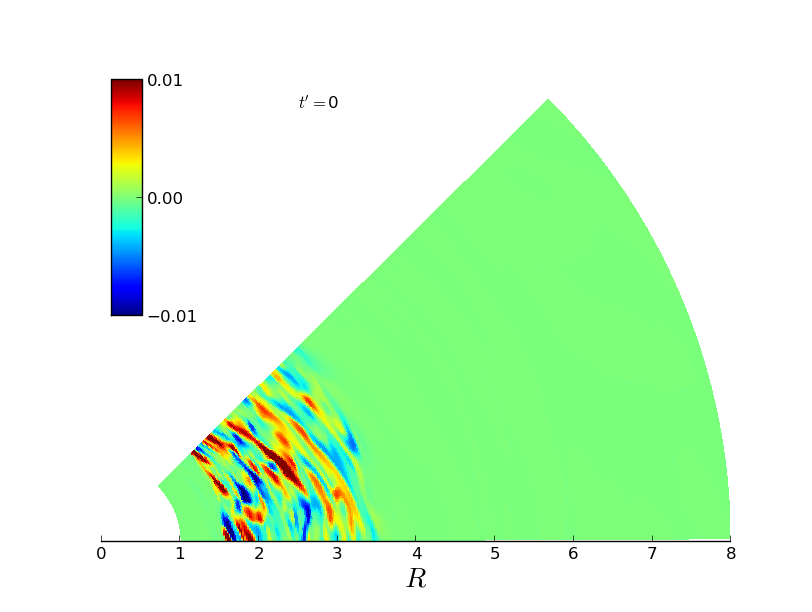}
\includegraphics[scale=0.28]{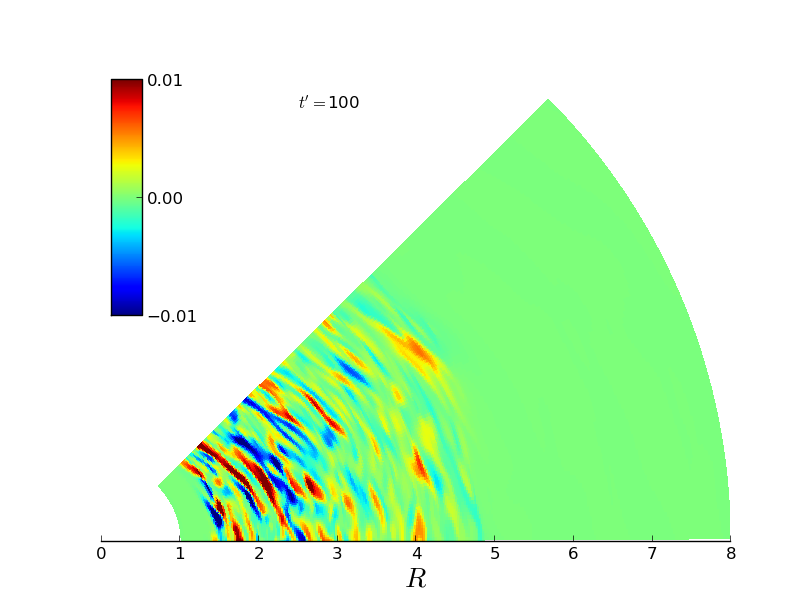}
\includegraphics[scale=0.28]{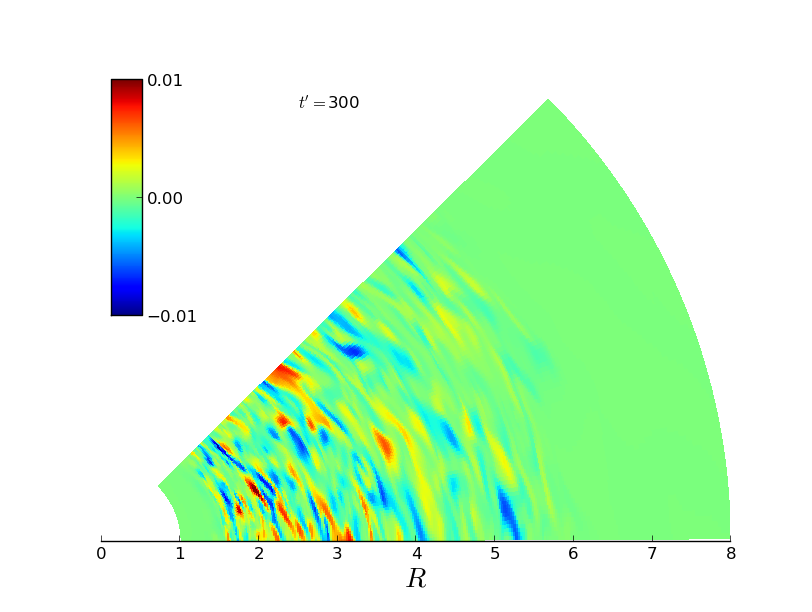}
\caption{Maps of radial magnetic field at different time $t=1600+t^{\prime}$. $t'$ is given in inner orbital time.} 
\label{frontMHD}
\end{center}
\end{figure*}

\begin{figure*}[!ht]
\begin{center}
\includegraphics[scale=0.37]{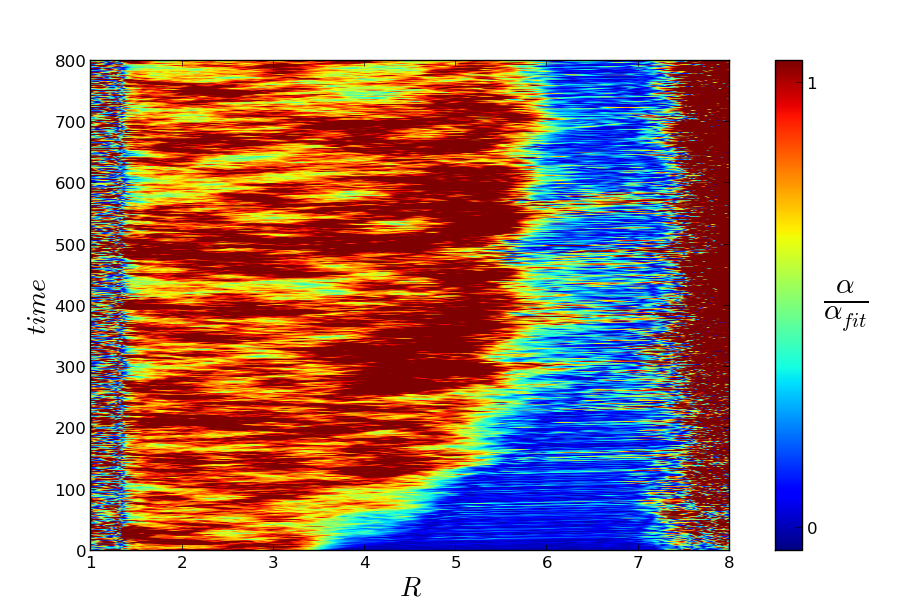}
\includegraphics[scale=0.37]{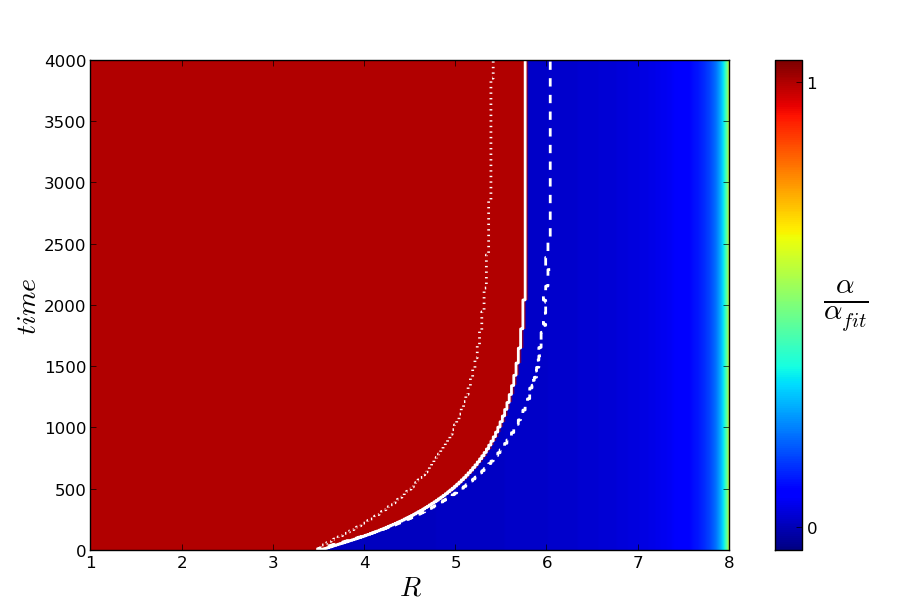}
\caption{Spacetime diagrams showing the turbulent activity
    evolution in the MHD simulation (left panel) and in the mean field
    model (right panel). In the case of the MHD simulation, $\alpha$
    is averaged over $Z$ and $\phi$. In the case of the mean
    field model, the different white lines mark
    the front position for the additional models described in
    section~\ref{MRIfront}. The solid line represents the case
    of remnant heating in the dead zone, the dashed line no remnant heating,
    and the dotted line the case of constant thermal diffusivity over the
    whole domain. Note the different vertical scales of
    the two panels (see discussion in the text for its origin).} 
\label{frontis}
\end{center}
\end{figure*}

We now move to the main motivation of this paper. At $t=1600$ we
restart model $\sigma_{hot}$ but close the feedback loop between
turbulence and temperature. The resistivity $\eta$ is now given by
Eq.~(\ref{resist}) with $\eta_0=10^{-3}$. We set the temperature
threshold to the value $T_\text{MRI}=0.4$, i.e. slightly above the typical 
temperature in the dead zone (which is about $0.25$ as discussed
above). 
This set-up ensures that at least half of the radial domain is
`bistable', i.e. gas at the same radius
 can support either one of two quasi-steady stable states, a
laminar cold state, and a turbulent hot state.
%% It is its behaviour in this
%% context that we test in this section. 

 As seen in figure~\ref{tempDZ}, the temperature exceeds
$T_\text{MRI}$ in the region $3.5 \le R \le 4.5$ at restart. Thus, we would
naively expect the interface to move to $R \sim 4.5$. As shown by
snapshots of the radial magnetic field fluctuations in the $(R,\phi)$
plane at three different times on figure~\ref{frontMHD}, indeed the
front moves outwards. In doing so, it retains its coherence.
However, as shown by the third panel
of figure~\ref{frontMHD}, the turbulent front
does not stop at $R \sim 4.5$ but moves outward all the way to $R \sim
5.5$--$6$. This is confirmed by the left panel of
figure~\ref{frontis}, in which a space-time diagram of $\alpha$
indicates that the front reaches its stagnation radius in a few hundred
orbital periods. 
\footnote{
To check that Eq.~(\ref{resist}) is an acceptable simplification of
the actual exponential dependence of $\eta$ with temperature, we ran
an additional simulation, with the same parameters as for the
$\sigma_{hot}$ but using Eq.~(\ref{eta_T}) for $\eta$. We find no
difference with our fiducial model: the front propagates equally fast
and stops at $R \simeq 5.3$, i.e. almost the same radius as for the
$\sigma_{hot}$ case.}
The following questions arise: can we predict the
stalling radius? And can we understand this typical timescale? What
can it tell us about the physical mechanism of front propagation?

\paragraph{Stalling radius:}
we use the mean field model proposed by \citet{latter&balbus12} to
interpret our results. The
front-stalling radius can be calculated from the requirement that the
integrated cooling and heating across the interface balance each other:
\begin{equation}
\int_{T_{DZ}}^{T_A}(Q^+-{\cal L})\,dT=0
\end{equation}
in which $T_{DZ}$ and $T_A$ stand for the temperature in
the dead zone and in the active zone respectively. 
While the cooling part of the integral is
well defined, the heating part $Q^+$ is more uncertain. It is
probably a mixture of turbulent heating (with an effective $\alpha$
parameter that varies across the front) and wave heating such as described
in section~\ref{static_interface_sec}. For simplicity, we adopt the
most naive approach possible and assume that $Q^+$ is a constant
within the front (equal to $1.5 \Omega T_{R\phi}$) as long as the
temperature $T$ is larger than $T_{MRI}$, and vanishes otherwise. In
that situation, we obtain, similarly to \citet{latter&balbus12} that
the stalling radius $R_c$ is determined by the implicit relation:
\begin{equation}
\label{criteria}
T_A(R_c)=\frac{5}{4}T_\text{MRI} \, .
\end{equation}
In figure \ref{tempDZ}, we draw two horizontal lines that correspond
to $T=T_{MRI}$ and $T=(5/4) T_{MRI}$. The condition given by
Eq.~(\ref{criteria}) is satisfied for $R \sim 5.25$ which is very
close to the critical radius actually reached in simulations (see
figures~\ref{frontMHD} and the left panel of
figure~\ref{frontis}). 

\begin{figure*}[!ht]
\begin{center}
\includegraphics[scale=0.23]{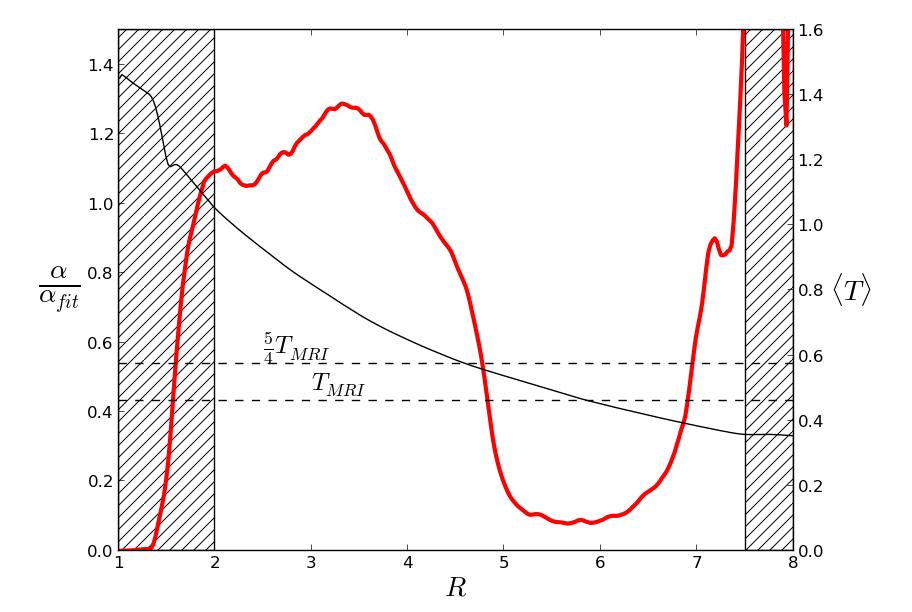}
\includegraphics[scale=0.23]{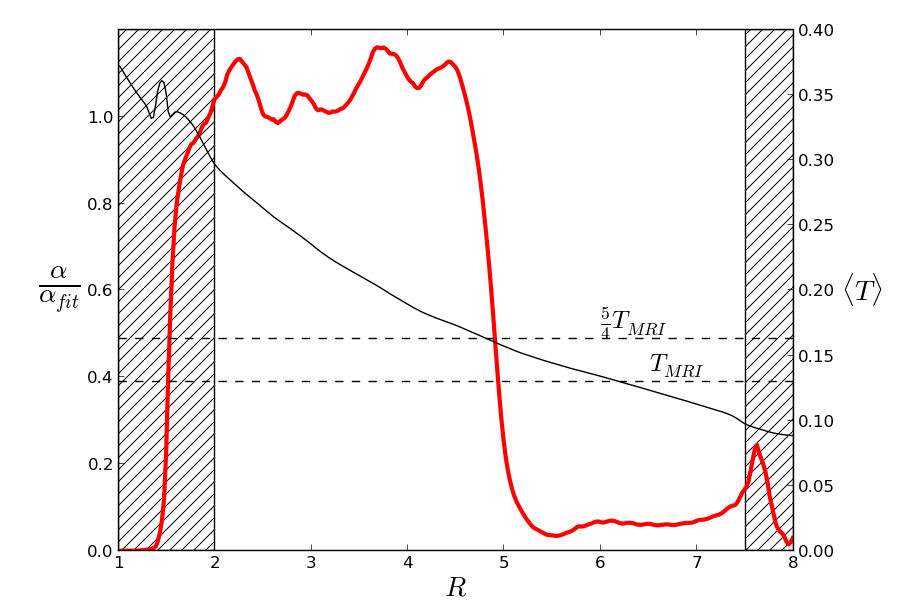}
\includegraphics[scale=0.23]{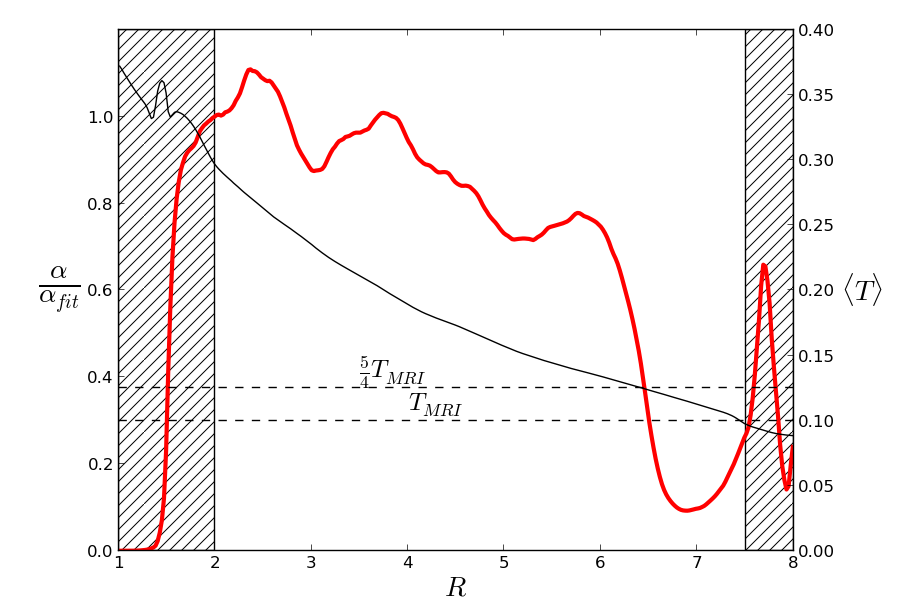}
\caption{$\alpha$ profiles averaged over 100 orbits after the fronts
  have reached their final position ($\alpha$ values are listed on the
  left axis). Left panel: $\sigma_{hot}$ simulation which $T_{MRI}=0.46$.
  Middle and right panel: $\sigma_{cold}$ simulation which
  $T_{MRI}=0.1$ and $T_{MRI}=0.13$ respectively.
  The black plain lines remind the temperature profiles from ideal
  cases. Black dashed lines show the threshold value $T_{MRI}$ and
  $5/4 T_{MRI}$. The temperature values are listed on the right axis} 
\label{finalfantasy}
\end{center}
\end{figure*}

In order to investigate the robustness of this result, we have carried
a further series of numerical experiments. First, we have repeated model
$\sigma_{hot}$ using a value of $T_{MRI}=0.46$ (instead of
$T_{MRI}=0.4$). According to
figure~\ref{tempDZ}, we expect the interface now to propagate over a
smaller distance. As shown by the left hand side panel of
figure~\ref{finalfantasy}, this is indeed the case: the front stalls
at $R \sim 5$ which is precisely the position where $T=(5/4)
T_{MRI}$.

We also computed two analogues of those models with $\sigma_{cold}$. 
This was done as follows: after the
$\sigma_{cold}$ ideal MHD simulations has reached a quasi thermal
equilibrium ($t=900$), we introduced a static dead zone from $R=3.5$ to
the outer boundary of the domain and we restarted this simulation. We
waited for thermal equilibrium to close the feedback loop at
$t=1600$. We used $T_{MRI}=0.1$. A front propagates until the critical
radius $R \sim 6.5$. It is as fast as the front of the $\sigma_{hot}$ case.
We show in the middle panel of figure~\ref{finalfantasy} that, in
accordance with the previous result, the front stops close to the
critical radius predicted by Eq.~(\ref{criteria}). We repeated the same
experiment using $T_{MRI}=0.13$ and show on the right panel of
figure~\ref{finalfantasy} that the front stalling radius is again
accounted for by the same equation.

To summarize, the front stops close to the predicted value in every
simulations we performed. This shows that the equilibrium position of
a dead zone inner edge can be predicted as a function of the disk's
radiative properties and thermal structure. 
It is robust, and in particular does not
depend on the details of the turbulent saturation in our simulations
(such as the radial $\alpha$ profile). 

%{\bf HENRIK: SHOULD WE MOVE THIS TO THE DISCUSSION (OR REMOVE IT?) --
%  SEE POINT 30 OF THE REFEREE --
%The general picture that describes both the mean field model and the
%simulations revolves around the relative basins of attraction of the
%two available states: the cold dead state and the hot active state.
%In the region of front propagation, either state is
%available but the front will move towards a radius 
%at which point the basins of attraction of both the
% dead state and the active state are of equal size. When these are
% unequal, perturbed gas near the interface will more easily migrate to the more
% attracting state and in so doing push the front along (see figure~5
% in Latter \& Balbus 2012). And because of the turbulence (and density
% waves) there is a constant supply of heating and cooling fluctuations near
%the interface that could potentially push gas from one basin to
%another.}

\paragraph{Front speed and propagation: }

the previous result is based on dynamical systems arguments. A front
is static at a given radial location if the attraction of the active turbulent
state balances the attraction of the quiescent state. 
However, the argument does not help us identify the front propagation
speed. In order to understand its dynamics, 
we ran a set of mean field simulations akin to the `slaved model' in
\citet{latter&balbus12} but using parameters chosen to be as
close as possible to model $\sigma_{hot}$. We solve
the partial differential equation for the temperature (we have
dropped the overbar here for clarity):
\begin{equation}
\frac{\partial T}{\partial t}=1.5 \Omega
T_{R\phi}-\frac{\gamma(\gamma-1)\sigma}{c_0^2}(T^{4}-T_{min}^4)
+\frac{\kappa_T}{R}\frac{\partial}{\partial R}\left(R\frac{\partial
  T}{\partial R}\right) \, .
\label{eq2}
\end{equation}
To solve that equation, we have used the heating term measured
during the ideal MHD simulation $\sigma_{hot}$ (see
section~\ref{ideal_mhd_sec}). The thermal transport of energy is
modeled by a simple diffusive law that is supposed to account for
turbulent transport. As discussed in
section~\ref{static_interface_sec}, the dead zone is
heated by waves excited at the dead/active interface. It is thus
highly uncertain (and one of the purpose of this comparison) whether a
diffusive model adequately describes heat transport of this fashion. 
The value for
$\kappa_T$ is chosen so heat diffusion
 gives the same mean flux of thermal energy as
measured in our simulation (shown in figure~\ref{divDZ}). We have
simplified the radial profile in that plot and have assumed
that the thermal flux vanishes outward of the percolation region but
is uniform in the active zone and in the percolation region: 
\begin{equation}
\label{diffus}
\kappa=
\begin{cases}
  \kappa_0 \, \, \, \, \textrm{if} \, \, \, R<R_f(t)+L_p\\
  0 \, \, \, \, \, \textrm{otherwise}.
\end{cases}
\end{equation}
The front location $R_f(t)$ is evaluated at each time step. It is the
smallest radius where $T<T_{MRI}$. As in the MHD simulation, we used
$T_{MRI}=0.4$ and $T_{min}=0.05$. When $L_p=0$, we have found the dead
zone inner edge is static regardless of the value of $\kappa_0$. As
expected, the front displacement requires the active and the dead zone
to be thermally connected. We have thus used a
percolation length $L_p=0.5$ and a
value of $\kappa_0=2.5 \times 10^{-4}$ in the active part of the disk which
matches the thermal flux at the outer edge of the percolation region
(see figure~\ref{divDZ}). The simulation is initialized with
the same physical configuration as the MHD simulation: a dead zone
extends from $R=3.5$ to $R=8$ and is in  thermal equilibrium. As shown
on the right panel of figure~\ref{frontis}, we find that a front
propagates outward. The \textit{critical radius} $R_c \simeq 5.5-6$ is
in agreement with the results of the simulation and with the argument 
based on energy conservation detailed above. However, there is clearly
a difference in the typical timescale of the propagation: the front
observed in the MHD simulation propagates five time faster than the
front obtained in the mean field simulation.

In order to check the sensitivity of this results to some of the
uncertain aspects of the mean field model, we have run three
additional mean field simulations similar to that described above.
In these additional models, we change one parameter 
while keeping all others fixed. In the first model, we have
changed the minimum value of the temperature $T_{min}=0.3$ (this is a
crude way to model the effect of the dead zone heating at long
distances from the interface). In the second model, we have used
$\kappa_0=5 \times 10^{-4}$. In the last model, we have used $L_p=+\infty$,
thereby extending turbulent heat transport to the entire radial extent
of the simulation. The results are shown on the right panel of
figure~\ref{frontis}: in all cases, the front propagates outward
slower than observed in the MHD simulation.

 A slow front velocity is thus a
generic feature of the diffusive approximation to heat
transport. But conversely, it reveals that the fast fronts mediated by real MRI
turbulence are controlled by non-diffusive heat transport. This in turn
strongly suggests that the front moves forward via the action of
fast non-local density wave heating, and not via the slow local turbulent
motions of the MRI near the interface, as originally proposed by Latter
\& Balbus (2012).  
 The transport via density waves occurs at a
velocity of order $c_s$. Thus, it takes a time $\Delta t_{waves} \sim
H/c_s \simeq \tau_{orb}/2 \pi$ for thermal energy to be transported through
the percolation length $L_p \sim H$. This is shorter than the typical
diffusion time over the same distance $\Delta t_{\textrm{diff}}=H^2/\kappa >
100 / 2 \pi \tau_{orb}$ at $R=3.5$, for the values of $\kappa$ used in
Eq~(\ref{eq2}). 
%% \textbf{The radiative diffusion time $\Delta
%% t_{\textrm{rad}}$ is also much longer than $\Delta t_{\textrm{waves}}$. 
%% The MRI front is at least $\Delta_R=1$ away from the sublimation front, the radiative diffusion time
%% over the distance between the two fronts is then much larger than $\Delta t_{waves}$.
%% Hence, the significant variation of $\sigma$ across the sublimation threshold do not play an important role in the front dynamics.
%% These different estimates illuminate the different mechanisms at work in each model. 
%% } 

%%% Local Variables: 
%%% mode: latex
%%% TeX-master: "main"
%%% End: 

\section{Instability and structure formation}
\label{structure_formation_sec}

Other issues that can be explored through our simulations
 are instability and structure formation at the dead/active
zone interface and throughout the dead-zone. The
extremum in pressure at the interface is likely to give rise to a
vortex (or Rossby wave) instability
\citep{1999ApJ...513..805L,varniere&tagger06,meheutetal12}. 
On the other hand, the
interface will control both the midplane temperature and density structure
throughout the dead zone; it hence determines the magnitude of the
squared radial Brunt-V\"ais\"al\"a frequency $\langle N_R
\rangle$. The size and sign of 
this important disk property is key to the emergence of the
subcritical baroclinic instability and resistive double-diffusive
instability in the dead zone 
\citep{lesur&pap10,2010MNRAS.405.1831L,2003ApJ...582..869K,
  petersenetal07a,petersenetal07b}. In this final section we briefly
discuss these instabilities, leaving their detailed numerical analysis
for a future paper. 

\subsection{Rossby wave instability}
\label{rossby_wave_sec}

\begin{figure*}
\begin{center}
\includegraphics[width=7cm]{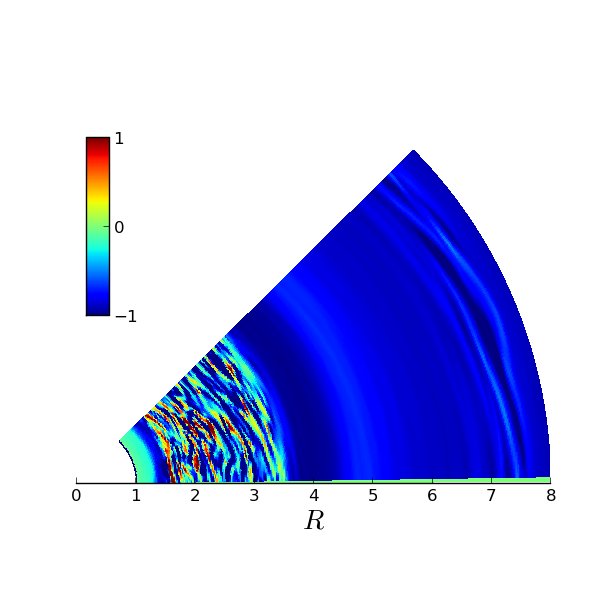}
\includegraphics[width=7cm]{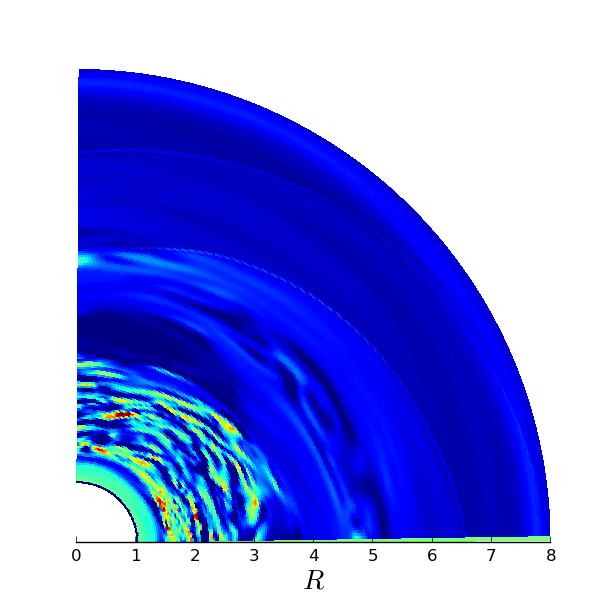}
\caption{Vortensity ($(V-V_{init})/V_{init}$) map in the disk
  $(R,\phi)$ plane. The
  left panel shows the midplane vortensity in the $\sigma_{hot}$
  simulation with the small azimuthal extend. The right panel shows the
  midplane vortensity in the $\sigma_{hot}$ simulation which has $\phi \in
  [0,\pi/2]$} 
\label{vort}
\end{center}
\end{figure*}

The Rossby wave instability has been studied recently by 
 \citet{lyra&maclow12} with MHD simulations that
use a locally isothermal equation of state and a static dead
zone. Vortex formation mediated by the Rossby wave instability 
was reported: a pressure bump forms at the interface
which then triggers the formation of a vortex. It is natural to wonder how
these results are modified when better account is made of the gas
thermodynamics.

In our simulations of a static dead zone
(section~\ref{static_interface_sec}), we also find that pressure
maxima form. Even with a $T$ dependent $\eta$, the bumps survive as
the interface travels to its stalling radius. Moreover, the amplitude of the
pressure bump is not modified compared to those observed in simulations
of static dead zones. On account of the reduced azimuthal extend of
our domains, no large-scale vortices formed. (The reduced azimuth
was chosen so as to minimize the computational cost of our
simulations.) 
In order to observe the development of the Rossby wave instability
 we performed one run identical to the
$\sigma_{hot}$ simulation except we extended the azimuthal domain:
$\phi=[0,\pi/2]$. The $\eta$ was a given function of position as in
Section 4.1, thus the dead/active zone interface was fixed. 
The right panel of figure~\ref{vort} shows a late snapshot of this
run, in which a vortex has appeared
similar to those observed by \citet{lyra&maclow12}. It survives for many dynamical
timescales. Consistent with the results of \citet{lyra&maclow12}, we
 measure $\alpha \simeq 0.01$ in the dead
 zone. This is two orders of magnitude larger than in the
 $\sigma_{hot}$ model. The dead zone is consequently much hotter in
 this simulation, which means the vortex plays a crucial role in
both accretion and the thermal physics of the dead
 zone. Simulations are currently underway to investigate the
 robustness of this results and the vortex survival when we close the
 feedback loop (setting $\eta=\eta(T)$). This will be the focus of a
future publication.

\subsection{Subcritical baroclinic and double-diffusive instabilities}

%% \begin{figure}
%% \begin{center}
%% \includegraphics[width=7cm]{./brunt.eps}
%% \caption{Mean profile of the square of Brunt-V\"ais\"al\"a frequency in the
%%   $\sigma_{hot}$ case including a static dead zone. The black line
%%   shows the mean profile of brunt-V\"ais\"al\"a frequency in the ideal
%%   $\sigma_{hot}$ case.} 
%% \label{brunt}
%% \end{center}
%% \end{figure}

Another potentially interesting feature of the interface is the strong
entropy gradient that might develop near the interface. It could also
impact on the stability of the flow on shorter scales, giving rise
potentially to the baroclinic instability
\citep{lesur&pap10,2013ApJ...765..115R} or to the double--diffusive 
instability \citep{2010MNRAS.405.1831L}. Both instabilities are
sensitive to the sign and magnitude of the entropy gradient, which is
best quantified by the Brunt-V\"ais\"al\"a frequency $N_R$: 
\begin{equation}
{\langle N_R \rangle}^2=-\frac{1}{\gamma \langle \rho \rangle}\frac{\partial \langle P \rangle}{\partial R}\frac{\partial}{\partial R}\ln{\frac{\langle P \rangle}{{\langle \rho \rangle}^{\gamma}}}.
\end{equation}
In general, we find that ${\langle N_R \rangle}^2$ takes positive values of
order $10$--$20 \%$ of the angular frequency squared. Negative values sometimes
appear localized next to the interface. Positive values
rule out both the baroclinic instability and the
double--diffusive instability, and indeed, we see neither in the
simulations. However, we caution against any premature conclusions about the prevalence 
of these instabilities in real disk. 
First, the source term in the continuity equation might
alter the density radial profile and, consequently, the
Brunt-V\"ais\"al\"a frequency in the dead zone. Second, it is well
known that these instabilities are sensitive to microscopic heat
diffusion. We do not include such a process explicitly in our
simulations. Instead, there is numerical diffusion of heat 
by the grid, the nature of which may be unphysical. Both reasons
preclude definite conclusions at this stage. Dedicated
and controlled simulations are needed to assess the existence and
nonlinear development
of these instabilities in the dead zone.

\section{Conclusion}
\label{conclusion_sec}

In this paper, we have performed non-ideal MHD simulations relaxing
the locally isothermal equation of state commonly used. 
We have shown the active zone strongly influences the thermodynamic
structure of the dead zone via density waves generated at the
interface. These waves transport thermal energy from the interface
deep into the dead zone, providing the dominant heating
source in its inner $20H$. As a consequence, the
temperature never reaches the very low level set by irradiation. In
the outer regions of the dead-zone, however, temperatures will be set 
by the starlight reprocessed by the disk's upper layers
\citep{chiang&goldreich97,dalessioetal98}. It is because the wave
generation and dissipation is located at the midplane that waves
should so strongly influence the thermodynamic structure of the
dead zone. 
%Heat sources from the warm upper layers are inefficient
%at transporting energy towards the midplane because such transport must battle
%a stable vertical entropy gradient.
Note also that density waves
generated by the dead/active zone edge are stronger than those excited
by the warm turbulent upper layers of the disk as seen in
  stratified shearing box \citep{fleming&stone03}.

Another result of this paper concerns the dynamical behaviour of the
dead-zone inner edge. We find the active/dead interface
propagates over several $H$ (i.e. a few  tens of an AU) in a few
hundreds orbits.All the simulated MRI fronts
reached a final position that matched the prediction made by a mean
field approach \citep{latter&balbus12}, which appeals to dynamical
systems arguments. As the gas here is bistable, it can fall into
either a dead or active state; the front stalls at the location 
where the nonlinear attraction of the active and dead states are in balance. 
In contrast, the mean-field model fails to correctly 
predict the velocity of the simulated fronts.
We find that a diffusive description of the radial energy flux yields
front speeds that are too slow. In fact, the simulations show that fronts
move rapidly via the efficient transport of energy by density waves across the
interface. Fronts do not propagate via the action of the slower
MRI-turbulent motions.

In addition, we have used our simulations to probe the
thermal properties of turbulent PP disks. 
We have constrained the turbulent
Prandtl number of the flow to be of order unity. We have also quantified
the turbulent fluctuations of temperature: they are typically of order a few
percent of the local temperature. However, their origin -- adiabatic
compression vs.\ reconnection -- is difficult to assess using
global simulations. In-depth dedicated non-isothermal shearing-box 
simulations will help to distinguish the dominant cause of the 
temperature fluctuations. Finally, we have made a first attempt to
estimate the radial profile of the radial 
Brunt-V\"ais\"al\"a frequency $N_R$ in the dead zone. This quantity is the key
ingredient for the development of both the subcritical baroclinic
instability and the resistive double-diffusive instability. We find
that $N_R$ can take both positive and negative values at different
radii; but we caution that these preliminary results require more testing 
with dedicated simulations.

Several improvements are possible and are the basis of future work. 
As discussed 
in section~\ref{structure_formation_sec}, one obvious extension 
is to investigate the fate and properties of emergent vortices
at the dead-zone inner edge. 
This can be undertaken with computational domains of a wider azimuthal
extent. Such simulations may investigate the role of large-scale
vortices, and the waves they generate, on the thermodynamic structure
of the dead-zone. They can also observe any feedback of the
thermodynamics on vortex production and evolution. 
We also plan
to investigate other magnetic 
field configurations. For example, vertical magnetic fields might
disturb the picture presented here because of the vigorous channel
modes that might develop in the marginal gas at the dead-zone
edge (Latter, Fromang, \& Gressel, 2010). 
Such an environment may militate against the development of
pressure bumps and/or vortices.   

Our results, employing the cylindrical approximation, represents
a thin
region around the PP disk midplane. These results must be extended so
that the vertical structure of the disk is incorporated. An urgent
question to be addressed is the location of density wave
dissipation in such global models. Waves can refract in thermally
stratified disks and deposit their energy at the midplane or in the
upper layers depending on the type of wave and the stratification
(Bate et al.~2002, and references therein). In particular, Bate et
al.\ demonstrate that large-scale 
axisymetric (and low $m$) density waves (f$^\text{e}$ modes) propagate upwards, as
well as radially, until they reach the upper layers of the disk, at which
point they transform into surface gravity waves and propagate along the disk surface.
But this is only shown for vertically polytropic disks in which the
temperature decreases with $z$ and for waves with relatively low
initial amplitudes. In PP-disk dead zones we expect the opposite to be
the case, and it is uncertain how density waves behave in this
different environment.

Finally, in this work we have increased the realism of one aspect of
the physical problem, the dynamics of the turbulence,
(via direct MHD simulations of the MRI), but have greatly 
simplified the physics of radiative cooling. As a result, the simulations
presented here are still highly idealized and several improvements
should be the focus of future investigations. For example, our approach
completely neglects the fact that dust sublimates when the temperature
exceeds $\sim 1500$~K. As a result, the opacity drops by up to four orders
of magnitude with potentially dramatic consequences for the disk
energy budget (leading to a increase of the cooling rate $Q_-$ of
  the same order, as opposed to our assumption of a constant
  $\sigma$). The radial temperature profile of our simulations 
indicates that the sublimation radius should be located $3$--$5$ disk
scaleheights away from the turbulent front. Given that the dominant
dynamical process we describe here is mediated by density waves
characterized by a fast timescale (compared to the turbulent and
radiative timescales), we do not expect the front dynamics
to be completely altered. Nevertheless, the relative proximity between
the sublimation radius and the turbulent front is still likely to
introduce quantitative changes. Clearly, the
thermodynamics of that region is more intricate than the simple
idealized treatment we use in this work. This further highlights
the need, in future work, for a more realistic treatment of radiative
cooling (for example using the flux limited diffusion approximation
with appropriate opacities, in combination with vertical
stratification. Such simulations will supersede our heuristic cooling
law with its constant $\sigma$ parameter. They are an enormous
challenge, but will be essential to test the robustness of the basic
results we present here.

\section*{ACKNOWLEDGMENTS}
JF and SF acknowledge funding from the European Research Council
under the European Union's Seventh Framework Programme (FP7/2007-2013)
/ ERC Grant agreement n° 258729. HL acknowledge support via STFC grant 
ST/G002584/1. The simulations presented in this paper were granted
access to the HPC resources of Cines under the allocation x2012042231
and x2013042231 made by GENCI (Grand Equipement National de Calcul
Intensif).

%%% Local Variables: 
%%% mode: latex
%%% TeX-master: "main"
%%% End: 

\begin{appendix}

\section{Opacity law in the buffer zones}
\label{buff_appendix}

Here, we give the functional form of $\sigma$ we used to
  prevent the temperature to drop at the inner edge of the simulation.
\begin{equation}
\sigma=
\begin{cases}
  \sigma_0\left(1-\frac{R}{2R_0}\right)^{-4} \, \, \, \, \textrm{if} \, \, \, R<2 R_0\\
  \sigma_0 \, \, \, \, \, \textrm{otherwise},
\end{cases} 
\end{equation}
where $\sigma_0$ stands for $\sigma_{\text{hot}}$ or
$\sigma_{\text{cold}}$ depending on the model. The value of $\sigma$ is kept
constant in the outer buffer.

\section{Wave heating}
\label{wave_heating_ap}

\begin{figure}
\begin{center}
\includegraphics[width=7cm]{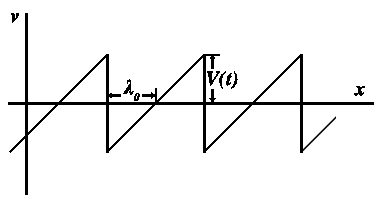}
\caption{The velocity fluctuations profile: Series of ``teeth'' modelling the wave shocks.} 
\label{zigzag}
\end{center}
\end{figure}

The large-scale density waves witnessed in our simulations
  develop weak-shock profiles, which are controlled by a competition
  between nonlinear steepening and wave dispersion. Keplerian shear
  may also play a role as it `winds up' the spiral and decreases the
  radial wavelength; though by the time this is important most of the
  wave energy has already dissipated.

A crude model that omits the strong dispersion inherent in our
  large-scale density waves nevertheless can successfully
  account for the energy dissipation in the simulations.
In such a model the density wave profiles are dominated by
  steepening and can thus be approximated by a sawtooth shape
  propagating at the sound speed velocity
$c_s$ (see figure~\ref{zigzag}). The evolution of the amplitude of
such isentropic waves is 
given by \citet{1959flme.book.....L}. In the wave frame of reference,
the gas velocity at the shock crest evolves over time as the shock
wave dissipates: 
\begin{equation}
v(t)=\frac{v_0}{1+\frac{(\gamma+1)}{2\lambda_0} v_0 t} \, ,
\end{equation}
where $v_0$ is the excitation amplitude of the wave and $\lambda_0$ its
wavelength, assumed to be conserved over the wave propagation. The
mean mechanical energy embodied in one wave period at time $t$ is
given by an integral over radius:
\begin{equation}
E_t=\frac{\rho}{\lambda_0} \int_{-\lambda_0/2}^{\lambda_0/2} \left(
\frac{R}{\lambda/2} - 1 \right)^2 v(t)^2 dR = \frac{1}{12}\rho v(t)^2 
\label{nrj_eq}
\end{equation}
where $\lambda$ is the wavelength at time $t$. We next compute the
mechanical energy radial flux through a unit surface:
\begin{equation}
F_R=E_t c_s  = \frac{\rho v(t)^2 c_s}{12}
= \frac{\gamma P_0 c_s}{12}  \left( \frac{\delta \rho}{\rho}\right)^2  
\label{flux_eq}
\end{equation}
where $\delta \rho$ is the difference between the shocked and
pre-shocked density. To compute the last equality, we have used the
fact that, under the weak shock approximation, the wave evolution
is isentropic and thus $\delta \rho / \rho= v/c_s$
\citep{1959flme.book.....L}. The wave energy and its flux are related
through the following conservation law:
\begin{equation}
\frac{\partial E_t}{\partial t}+\frac{1}{R}\frac{\partial R
  F_R}{\partial R}=0 \, ,
\label{conserve_eq}
\end{equation}
while the dissipation rate of the wave is expressed using the mechanical energy
conversion into heat per unit time:
\begin{equation}
D=\frac{\partial E_t}{\partial t}=\frac{1}{12} \rho v \frac{\partial v}{\partial t} = \frac{\gamma (\gamma+1)}{12} P_0 \left(\frac{\delta \rho}{\rho}\right)^3 f
\end{equation}
where $f=c_s/\lambda_0$ is the wave frequency. We used the density
fluctuations in our simulations as the difference between the shocked
and pre-shocked density to estimate the local wave heating in
Eq.~(\ref{shoki}).

\noindent
In addition to that estimate, the thermal energy flux divergence can
be used, through Eq.~(\ref{conserve_eq}), as a way to estimate the
radial variation of the wave amplitude: 
\begin{eqnarray}
\frac{1}{R}\frac{\partial R F_R}{\partial R} &=& \frac{1}{R}\frac{\partial }{\partial
  R}\left[\frac{\gamma P_0 c_s R}{12}  \left( \frac{\delta
  \rho}{\rho}\right)^2 \right] \\
&=& \frac{\gamma P_0 c_s}{12} \left[ \left(
  \frac{\delta \rho}{\rho}\right)^2 \frac{1}{P_0}\frac{\partial P_0}{\partial
    R}+ \left( \frac{\delta
    \rho}{\rho}\right)^2 \frac{1}{c_s}\frac{\partial c_s}{\partial R}
  \right. \nonumber \\
&+& \left. 2\left( \frac{\delta \rho}{\rho}\right)\frac{\partial }{\partial
    R} \left( \frac{\delta \rho}{\rho}\right)+ \frac{1}{R} \left( \frac{\delta
    \rho}{\rho}^2\right) \protect\right] \, . \nonumber
\end{eqnarray}
This must equal the energy rate released as thermal heat given by
$-D$. Combining the last two expression thus provides an expression
for the radial decay of the wave amplitude:
\begin{equation}
\frac{\partial }{\partial R} \left( \frac{\delta \rho}{\rho}\right) =
\left( \frac{\delta \rho}{2 \rho}\right) \left(\frac{1}{R} -
\frac{1}{P_0}\frac{\partial P_0}{\partial R} -
\frac{1}{P_0}\frac{\partial c_s}{\partial R} - \frac{\gamma+1}{c_s}
\frac{\delta \rho}{\rho} f\right) \, .
\label{wavedamp}
\end{equation}
The first term is the geometrical term that described the wave
dilution as it propagates cylindrically. The second and third terms
are specific to waves propagating in stratified media where mean
pressure and sound speed are not uniform. Finally, the last term of
the right hand side of this equation reflects the wave damping by
shocks. In our simulations, all four terms are of comparable
importance. 

\section{Model with an extended radial extent}
\label{large_radius_app}

\begin{figure}
\begin{center}
\includegraphics[width=7cm]{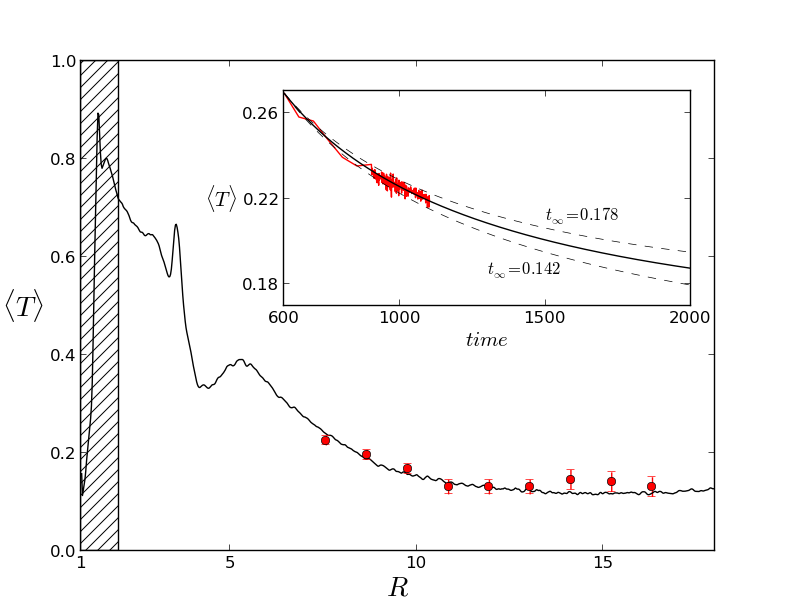}
\caption{Temperature profile $\langle T \rangle$ averaged over $200$
  orbits after $t=900+\tau_{cool}(R=7)$ obtained in the extended
  $\sigma_{hot}$ case. The black line show wave-equilibrium
  temperature profile. Red dots show the extrapolated temperature for
  $7$ radii. The inserted frame show in red the temperature evolution
  $\overline{T}$ after $t=900$ at $R=9.8 R_0$ in the extended
  $\sigma_{hot}$ simulation. On this subplot, the black plain line
  shows the accurate fit and the black dashed lines show the two
  extreme fits used to determine the measurement error.} 
\label{nrjwav}
\end{center}
\end{figure}

In this section of the appendix we describe the results obtained
 in the radially extended $\sigma_{hot}$ run.
We initiated MHD turbulence in that model in the absence of
any dissipative term until the region between $R=1$ and $R=3.5$ has
reached thermal equilibrium ($t=900$). At that point, we set
$\eta_0=10^{-3}$ for $R \ge 3.5$. As expected, a dead zone quickly
appeared at those radii. Because of the prohibitic computational coast
of that simulation (there are $960$ cells in the radial direction!),
we were not able to run that model until thermal equilibrium is
established at all radii. Indeed, the cooling time becomes very long
at large radius. Instead, we followed the time evolution of the
temperature at nine locations in the dead zone. Due to the absence of
turbulent heating, we found that the temperature slowly decreases with
time. Assuming this decrease is due to a combination of cooling
(resulting from the cooling function) and wave heating, it
can be modelled using the following differential equation
\begin{equation}
\frac{\partial T}{\partial t}=-\frac{\gamma (\gamma-1)\sigma_{hot}}{c_0^2} T^4 + Q_{w}^+.
\end{equation}
where the term $Q_w^+$ accounts for the (unknown) wave heating. We
fitted the time evolution of the temperature during the duration of
the simulation ($\sim 1000$ orbits) in order to obtain a numerical estimate
of $Q_w^+$ at those nine locations. An example of that fit is
provided in the insert of figure~\ref{nrjwav}. Using these value, we
can obtain an estimate for the equilibrium temperature in the dead
zone, as shown on figure~\ref{nrjwav}. The comparison
with the estimate of Eq.~(\ref{temp_dz_eq}) provided by the black line
is excellent everywhere in the dead zone.

\begin{figure}
\begin{center}
\includegraphics[width=7cm]{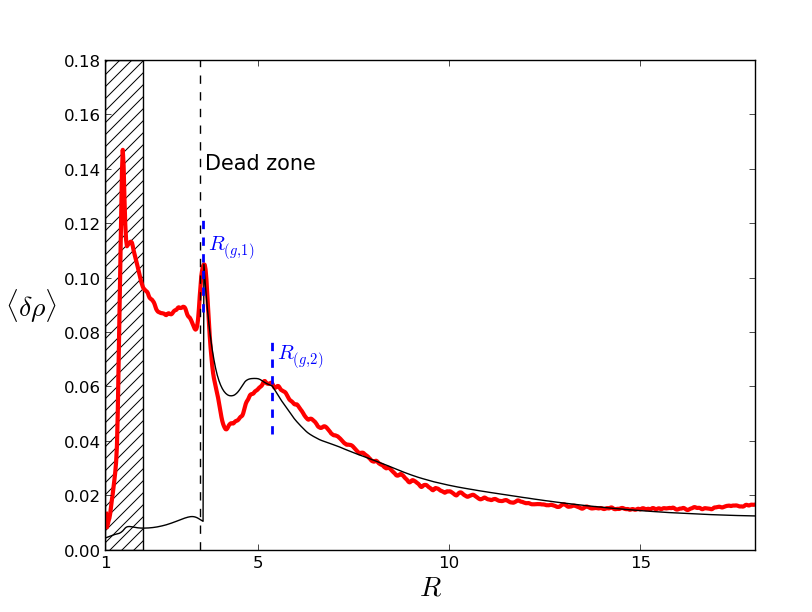}
\caption{Profile of density standard deviation $\langle \delta \rho
  \rangle$ averaged over $200$ inner orbits after
  $t=900+\tau_{cool}(R=7)$ in the extended $\sigma_{hot}$ case. The
  black plain line shows the density fluctuation amplitude deducted
  from the two waves model. The excitation locations used in the two
  waves model are shown by blue dashed lines.} 
\label{damp}
\end{center}
\end{figure}

As a sanity check, we test here if the wave amplitude
decreases as their energy content is converted into heat.
We show on figure~\ref{damp} the mean density fluctuation profile
obtained in the extended $\sigma_{hot}$ simulation. 
It exhibits two maxima close to the dead zone inner edge located at
$R_{(g,1)}$ and $R_{(g,2)}$ and the amplitude decay with radius. The
reason why we see two such maxima is not clear but might be due to
waves originated at the dead/active interface as well as waves excited
as the location of the pressure maximum. In any case, we found that
modelling the amplitude of fluctuations as the signature of a 
combination of two waves generated at $R_{(g,i)}$ gives acceptable
results. We use an explicit scheme to numerically integrate
Eq.(\ref{wavedamp}) from the wave generation locations $R_{(g,i)}$. 
The two waves are excited with the amplitude measured at $R=R_{(g,i)}$
with the frequency $1/f=0.15\tau_{cool}(R_{(g,i)})$. We plot the
solution thus obtained on figure~\ref{damp}. The good agreement between the
analytical solution and the profile gives a final confirmation that
waves control the dead zone thermodynamics.

\section{Simulation with a higher resolution}
\label{high_res_sim}
\begin{figure}
\begin{center}
\includegraphics[width=7cm]{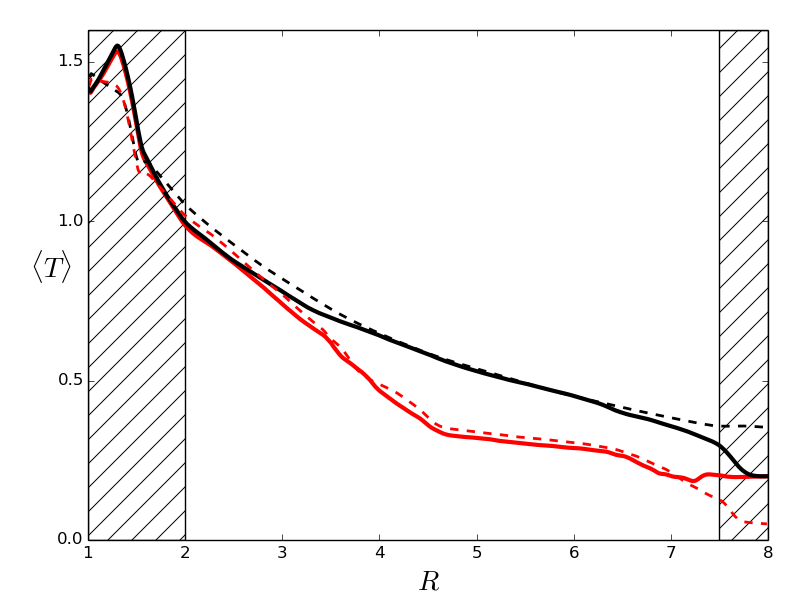}
\caption{Profiles of temperature $\langle T \rangle$ averaged over $100$ inner orbits after
  $t=700$ and $t=1000$ in the highly resolved $\sigma_{hot}$ case (plain lines).
  The dashed lines remind the temperature profiles obtained with the low resolution run.
  For both resolutions, the temperature in the ideal case is shown by a red line 
  and the temperature in the static dead zone case is shown by a black line.}
\label{high_res_T}
\end{center}
\end{figure}

Here we present a brief test of the impact of spatial resolution on our results.
We have restarted the $\sigma_{hot}$ run from the thermal equilibrium
of the ideal MHD case ($t=600$) and the static dead zone case
($t=900$), with twice as many cells in each direction. The
resolution is $(640,160,160)$. We show the averaged temperature
profile of both cases on figure~\ref{high_res_T}.
Because of the large computational cost, each models are
integrated for $200$ orbits and time averages are only performed over
the last $100$ orbits. The temperature
profiles of each case are very close to those obtained with the fiducial runs.
We conclude that resolution as little impact on our results.

%% \section{Radiative flux}

%% In optically thick region, the radiative transfer is well approximated by a diffusive approach.
%% Assuming that the radiative diffusion is isotropic, the radiative flux writes:
%% \begin{equation}
%% \label{rad_flux}
%% \vec{F_{Rad}}=\frac{4}{3}\frac{\sigma_b T^3}{\rho k}\vec{\nabla}T
%% \end{equation}
%% with $\sigma_b$ the constant of Stefan-Boltzmann and $k$, the opacity.
%% This equation translates into the MHD simulations unit system of section~\ref{init_sec} via the cooling term ${\cal L}_{cool}$ which models the energy radiatively expelled through the disk surface.
%% To approximate the Eq.(\ref{rad_flux}), we consider that a single source located in the midplane releases the energy $E_{sc}\simeq {\cal L}_{cool} H$ lost by the column of gas above it.
%% This energy is vertically transported by radiative diffusion.
%% In a steady state:
%% \begin{equation}
%% E_{sc}=\frac{4}{3}\frac{\sigma_b T^3}{\rho k}\frac{\partial T}{\partial z}.
%% \end{equation}
%% The temperature typically varies on one scale height: $\partial T / \partial z \simeq T/H$.
%% Using ${\cal L}_{cool} \simeq \rho \sigma T^4$, one finally obtains the radial radiative flux:
%% \begin{equation}
%% \vec{F_R}=H^2 \rho \sigma T^3\frac{\partial T}{\partial R}.
%% \end{equation}

\end{appendix}

%%% Local Variables: 
%%% mode: latex
%%% TeX-master: "main"
%%% End: 

\bibliographystyle{aa}
\bibliography{main}

\end{document}